\documentclass[twocolumn,aps,showpacs,amsfonts,amsmath, amssymb,floatfix,superscriptaddress]{revtex4-1}
\usepackage[final]{graphicx}
\usepackage{color}
\usepackage{amssymb,amsmath}

\newcommand{\vct}[1]{\mathbf{#1}}

\newcommand{\tp}{\Psi}

\newcommand{\kk}{\vct{k}}
\newcommand{\qq}{\vct{q}}
\newcommand{\rr}{\vct{r}}
\newcommand{\vv}{\vct{v}}
\newcommand{\xx}{\vct{x}}
\newcommand{\eeta}{\boldsymbol \eta}
\newcommand{\ssigma}{\boldsymbol \sigma}

\newcommand{\be}{\begin{equation}}
\newcommand{\ee}{\end{equation}}

\allowdisplaybreaks

\DeclareSymbolFont{bbgreek}{U}{bbold}{m}{n}
\DeclareMathSymbol{\bbmu}{\mathbb}{bbgreek}{'26}
\DeclareMathSymbol{\bbeps}{\mathbb}{bbgreek}{'17}

\definecolor{darkblue}{rgb}{0,0,0.6}
\definecolor{darkred}{rgb}{0.6,0,0}
\usepackage[colorlinks=true,urlcolor=darkblue,citecolor=darkblue,linkcolor=darkred,hyperfootnotes=false]{hyperref}

\begin{document}

\title{Stresses in non-equilibrium fluids: Exact formulation and coarse grained theory}

\date{\today}

\author{Matthias Kr\"uger}
  \altaffiliation{Current Address: Institute for Theoretical Physics, Georg-August-Universit\"at G\"ottingen, 37073 G\"ottingen, Germany}
\affiliation{Max Planck Institute for Intelligent Systems, Heisenbergstr. 3, 70569 Stuttgart, Germany}
 \affiliation{4th Institute for Theoretical Physics, Universit\"at Stuttgart, Pfaffenwaldring 57, 70569 Stuttgart, Germany}

 \author{Alexandre Solon}
 \affiliation{Department of Physics, Massachusetts Institute of Technology, Cambridge, Massachusetts 02139, USA}

 \author{Vincent D\'emery}
\affiliation{Gulliver, CNRS, ESPCI Paris, PSL Research University, 10 rue Vauquelin, 75005 Paris, France}
\affiliation{Univ Lyon, ENS de Lyon, Univ Claude Bernard Lyon 1, CNRS, Laboratoire de Physique, F-69342 Lyon, France}

\author{Christian M. Rohwer}
 
 \affiliation{Max Planck Institute for Intelligent Systems, Heisenbergstr. 3, 70569 Stuttgart, Germany}
 \affiliation{4th Institute for Theoretical Physics, Universit\"at Stuttgart, Pfaffenwaldring 57, 70569 Stuttgart, Germany}

\author{David S. Dean}
\affiliation{Univ. Bordeaux and CNRS, Laboratoire Ondes et Mati\`ere d'Aquitaine (LOMA), UMR 5798, F-33400 Talence, France}

\begin{abstract} 
  Starting from the stochastic equation for the density operator, we
  formulate the exact (instantaneous) stress tensor for interacting
  Brownian particles, whose average value agrees with expressions
  derived previously. We analyze the relation between the stress
  tensor and forces on external potentials, and observe that,
  out of equilibrium, particle currents give rise to extra
  forces. Next, we derive the stress tensor for a Landau-Ginzburg
  theory in non-equilibrium situations, finding an expression analogous
  to that of the exact microscopic stress tensor, and discuss the computation of out-of-equilibrium (classical) Casimir forces. We use these
  relations to study the spatio-temporal correlations of the stress
  tensor in a Brownian fluid, which we derive exactly to leading order
  in the interaction potential strength. We observe that, after
  integration over time, the spatial correlations generally decay as
  power laws in space. These are expected to be of importance for
  driven confined systems. We also show that divergence-free parts of the stress tensor do not contribute in the Green-Kubo relation for the viscosity.
\end{abstract}

\bibliographystyle{plain}
\maketitle

\section{Introduction}
The physics of out-of-equilibrium fluids is fascinating in its
complexity and the great variety of phenomena they exhibit. This is in
part due to the many ways a system can be put out of equilibrium,
e.g., by a continuous external driving such as an applied
shear~\cite{dhont,larson} or a temperature gradient~\cite{Ortiz},
or by a (sudden) change of a control parameter like a temperature quench
producing a supercooled liquid (see e.g. Ref.~\cite{Debenedetti01}).

In general, describing the static properties of fluids in thermal
equilibrium is already a formidable task due to the inherent many-body
interactions~\cite{HansenMcDonald}. In that respect, density
functional theory (DFT) \cite{bob_advances,roth_review} has been
most successful in the study of static density profiles and
correlation functions. Concerning the dynamics (or non-equilibrium
situations), insight has been obtained in the dilute limit from
exactly solvable two-body models \cite{Andrade,
  brady_morris, Harris04,dhont}, e.g. from the Boltzmann or the
Smoluchowski equation. For multibody dynamics, approximate treatments
include mode coupling theory \cite{Fuchs02, Miyazaki02} or dynamical
density functional theory (DDFT) \cite{archer,Marconi99}, including power functional theory \cite{SchmidtBraderJCP_2013_power_func}.
Non-equilibrium molecular fluids have also been studied extensively by
computer simulations
\cite{Zhang2004SimPoisFl,Glavatskiy15,Glavatskiy15b,roy2016}.

While the above-mentioned approaches start from the microscopic
details of the fluid (at the particle level), a very different
approach starts from effective field theories (such as Landau-Ginzburg
Theory \cite{hohenberg, onukibook,kardarbook}), where microscopic
details are neglected in order to study only the large-scale phenomena, for
example in near-critical fluids. Non-equilibrium scenarios also yield
insight and challenges in this context: Several computations have been
concerned with critical Casimir forces away from equilibrium, i.e.,
for temperature quenches \cite{gambassi2006,
  gambassi2008EPJB,deangopinathan2010PRE}, moving objects
\cite{demery2010,hanke,gambassi2013prl}, or shear in confinement
\cite{Rohwer17b}.
Non-equilibrium fluctuations arising from conservation laws have also
been demonstrated to lead to Casimir forces in various setups
\cite{wadasasa2003,sotogranular2006,shaebaniwolf2012,kirkpatricksengers2013,aminovkardarkafri2015,kirkpatrick2015prl,kirkpatrick2016pre,Rohwer17,Rohwer17c}. 
Moreover,
the pressure and stresses exerted by active systems have attracted
growing attention for their unusual
properties~\cite{Solon15,bialke2015negative}. 
In all these studies,
the computation of forces and stresses in out-of-equilibrium
situations is non-trivial, and the form of the applicable stress tensor 
has been discussed (partly controversially) 
\cite{deangopinathan2009JStatMech,deangopinathan2010PRE,bitbolfournier2011forces}.

In this manuscript, we derive several expressions for the stress
tensor of a liquid. In particular, starting from the microscopic
dynamics of interacting Brownian particles, we compute the exact
stress tensor for any given (snapshot) density realization. Importantly, the stress tensor's average
agrees with the form obtained from the Smoluchowski
equation~\cite{irving_kirkwood, Kreuzer, AerovKrugerJCP2014}. Before
averaging, the fluctuating form is naturally well suited for use in
Green-Kubo fluctuation relations~\cite{Green,Kubo}. As a second step,
we derive a non-equilibrium version of the stress tensor for a
Landau-Ginzburg Hamiltonian and identify the various contributions,
which we relate to those found in the exact
microscopic version. Using the (non-equilibrium) stress tensor to compute
the force on an embedded object, the results are in agreement with
existing literature~\cite{deangopinathan2010PRE}.
As an application, we use the expression for the instantaneous stress
tensor to compute temporal and spatial stress correlations, this can be done exactly to leading order in the interaction potential $V$. We show
that, after integration over time, the stress-stress correlations
decay as power laws in space. These correlations have been of recent
interest~\cite{Maier17}, see also Ref.~\cite{Evans81}, and are expected to influence the flow in
confinement~\cite{aerov2015} or for inhomogeneous flow velocities \cite{Machta79}.
Finally, we use the Green-Kubo formula to deduce the viscosity of the liquid from the stress correlations.

The manuscript is organized as follows: Sec.~\ref{sec:H} introduces
the system under consideration and defines several observables of
interest. Sec.~\ref{sec:sigma} gives the exact instantaneous stress
tensor for the density operator, while Sec.~\ref{sec:sigmalandau}
analyzes the stresses in a standard local field theory (with only zeroth and first derivative terms). Sec.~\ref{sec:App}
provides an application: We compute the correlations of the stress
tensor.

\section{The System and Observables}\label{sec:H}

\begin{table}
\begin{ruledtabular}
\begin{tabular}{|p{2.3cm}|p{5.9cm}|}
\hline
Symbol &Meaning\\
\hline\hline\hline
$\rho(\vct{x},t)$&Density operator: $\rho(\vct{x},t)=\sum_\mu\delta(\vct{x}-\vct{x}_\mu(t))$\\
\hline
$\bar\rho(\vct{x})=\langle\rho(\vct{x})\rangle^{\rm eq}$&Mean density in equilibrium.\\
\hline
$\phi(\vct{x},t)$& Fluctuation of density about its equilibrium value, $\phi(\vct{x},t)=\rho(\vct{x},t)-\bar\rho(\vct{x})$.\\
\hline
$\left\langle \phi(\vct{x},t)\phi(\vct{x}',t')\right\rangle$&Time-dependent correlations of density fluctuations.\\
\hline
$\langle\rho(\vct{x},t)\rangle$&Mean density.\\
\hline
$\rho^{(2)}(\vct{x},\xx')$&Averaged two-body density: $\rho^{(2)}(\vct{x},\xx')=\left\langle \sum_{\mu\neq \nu}\delta(\vct{x}-\vct{x}_\mu) \delta(\xx'-\vct{x}_\nu)\right\rangle$.\\
\hline
$g(\vct{r})$&Pair correlation function in bulk: $g(\vct{r})=\frac{1}{N\langle \rho\rangle}\left\langle  \sum_{\nu\not= \mu}\delta(\vct{r} -\vct{x}_\nu+\vct{x}_\mu) \right\rangle$.\\
\hline
\end{tabular}
\end{ruledtabular}
\caption{
\label{table:1}
Observables relevant for this  manuscript. Note that the mean density,
the two body density, and the pair correlation can be evaluated both in
or out of equilibrium.
}

\end{table}

\subsection{System}\label{sec:S}
To investigate non-equilibrium liquids, we choose the well-studied and
experimentally relevant model system of overdamped spherical
Brownian particles.
This system has the advantage that, even if driven far from
equilibrium, the solvent stays equilibrated in any situation and acts
as a bath at the given temperature. This way, a well characterized
out-of-equilibrium state is obtained.

Using Brownian dynamics directly implies a
canonical or grand-canonical description, where the solvent acts as
a bath at the given temperature. We generally consider
systems for which canonical and grand canonical descriptions are
equivalent due to a very large (infinite) particle number (for example 
a semi-infinite system bound by a planar surface). 

The Brownian particles with positions at $\vct{x}_\mu$ are subject to a
potential $\tp(\{\vct{x}_\mu\})$ that includes pairwise interactions
(denoted by $V$) as well as an external potential (denoted $U$),
\begin{equation}
\tp(\{\vct{x}_\mu\}) = \sum_{\mu<\nu}V(\xx_\mu-\xx_\nu) + \sum_\mu U(\xx_\mu).
\label{def_Phi}
\end{equation}
Indices $\mu$ and $\nu$ run over all particles.
The thermal energy scale is denoted by $k_BT\equiv\beta^{-1}$, with
Boltzmann's constant $k_B$ and the (solvent-imposed) temperature
$T$. The bare diffusivity (in the absence of interactions) of the Brownian
particles is denoted by $D$. Each particle thus obeys the overdamped
Langevin equation

\begin{equation}
\frac{d {\bf x}_\mu}{d t} = D\beta {\bf F}_\mu  + \sqrt{2D}\boldsymbol{\xi}_\mu,\label{sde}
\end{equation}
where $\boldsymbol{\xi}_\mu$ is a Gaussian white noise with zero mean
and with correlations $\langle \xi_{\mu,i}(t) \xi_{\nu,j}(t')\rangle
=\delta_{\mu\nu}\delta_{ij}\delta(t-t')$, and ${\bf F}_\mu=-\nabla_{\xx_\mu}\tp$ is the force
acting on particle $\mu$ due to the potential $\tp$. (Throughout, $i$
and $j$ label spatial components, while Greek indices label particles). If $\tp=0$, each particle performs
isotropic Brownian motion.

\subsection{Observables -- mean and fluctuating -- in and out of equilibrium}\label{sec:Ob}
We summarize the important observables for this manuscript in Table~\ref{table:1}. The
basic quantity is the density operator~\cite{HansenMcDonald},
\begin{align}\label{eq:do}
\rho(\vct{x},t)=\sum_\mu\delta(\vct{x}-\vct{x}_\mu(t)).
\end{align}
It is the starting point for all considerations that follow. Averaged
over the equilibrium distribution, one
obtains the mean equilibrium density $\bar\rho$, defined as
\begin{align}
\bar\rho(\xx)\equiv\left\langle
\rho(\vct{x})\right\rangle^{\rm eq}=
\left\langle\sum_\mu\delta(\vct{x}-\vct{x}_\mu)\right\rangle^{\rm eq}.\label{eq:barrho}
\end{align}
Here we have introduced the {\it equilibrium} average
$\left\langle \dots\right\rangle^{\rm eq}$,  which, for the overdamped
system is exactly given by
\begin{align}\label{eq:av0} \left\langle
\dots\right\rangle^{\rm eq}=\frac{\int d\Gamma \dots e^{-
\beta\tp(\Gamma)}}{\int d\Gamma e^{- \beta\tp(\Gamma)}},
\end{align}
where $\Gamma\equiv\{{\vct{x}_\mu}\}$.
As noted above, for large systems the grand canonical average agrees
with the canonical one given here. We introduce the density fluctuation field $\phi(\xx,t)$, which quantifies the deviation of the density
operator from its equilibrium mean
\begin{align}
\phi(\vct{x},t)=\rho(\vct{x},t)-\bar\rho(\vct{x}).  \label{eq:phi}
\end{align}
Fluctuations can be characterized by their two-point correlation
function
\begin{align}\label{eq:Cc} C(\vct{x},\vct{x}',t,t')=\left\langle
\phi(\vct{x},t)\phi(\vct{x}',t')\right\rangle,
\end{align}
where we introduced the average $\left\langle \dots\right\rangle$ over
noise realizations given a (possibly non-equilibrium) ensemble of
initial conditions. The correlation function in Eq.~(\ref{eq:Cc}) is
thus well-defined in or out of equilibrium. In stationary state, $C$
is a function of $t-t'$ only~\cite{Risken}, and depends only on the relative coordinate 
$\vct{x}-\vct{x}'$ in homogeneous systems. Its spatial Fourier
transform, $\tilde C$ is the intermediate scattering function
\cite{dhont}.

Another important quantity (related to $C$) is the two-body density,
defined by \cite{HansenMcDonald}
\begin{align}\label{eq:rho2}
\rho^{(2)}(\vct{x},\xx',t)=\left\langle \sum_{\nu\neq \mu}\delta(\vct{x}-\vct{x}_\mu(t)) \delta(\xx'-\vct{x}_\nu(t))\right\rangle.
\end{align}
For bulk systems, $\rho^{(2)}$ depends only on one coordinate
$\vct{r}$~\cite{HansenMcDonald} and can be expressed via the pair
correlation function $g$,
\begin{align}
g(\vct{r})&=\frac{1}{\langle\rho\rangle^2}\rho^{(2)}(\vct{x}+\vct{r},\vct{x})\nonumber\\ 
&=\frac{1}{N\langle \rho\rangle}\left\langle \sum_{\mu\neq \nu}\delta(\vct{r}+\vct{x}_\mu -\vct{x}_\nu) \right\rangle.
\label{eq:gd}
\end{align}
Again, the above average for $g$ can be evaluated in or out of
equilibrium. One example for a non-equilibrium pair correlation
function is found in systems under shear, where $g$ is distorted
compared to the reference equilibrium case \cite{dhont}.

\section{Exact microscopic  stress tensor}
\label{sec:sigma}
In this section, we derive and discuss the exact microscopic stress
tensor (including its fluctuations) for the system of Brownian particles.
\subsection{Microscopic theory}
The stress tensor is related to forces in the system. These forces can
be read off directly on the equation of motion, which for the density
operator $\rho$ is given by~\cite{Dean96}
\begin{align}
\frac{\partial\rho}{\partial t} ({\bf x},t) 
= \nabla \cdot \left[ D\rho(\xx,t)\nabla \frac{\delta \beta {\cal E}}{\delta \rho({\bf x},t)} + \sqrt{2D\rho({\bf x},t)}{\eeta}({\bf x},t)\right].
\label{dk1}
\end{align}
Eq.~(\ref{dk1}), interpreted with the It\=o convention, is an exact
reformulation of Eq.~\eqref{sde}. The term $\eeta$ is a vectorial
Gaussian white noise field with zero mean and correlations
\begin{align}
\left\langle\eta_i(\vct{x},t)\eta_j(\xx',t')\right\rangle= \delta_{ij}\delta(\vct{x}-\xx') \delta(t-t').
\label{eq:FDT}
\end{align}
{$\cal E$} is the energy functional, which contains an ideal gas part,
a contribution from interactions via the inter-particle potential $V$,
and the external potential $U$,
\begin{align}\label{eq:pot}
{\cal E}[\rho(\xx)] & = k_BT\int d{\bf x} \rho({\bf x}) \ln(\rho({\bf x})) \notag\\&\quad+ \frac{1}{2}\int d{\bf x}d\xx' \rho({\bf x}) V({\bf x}-\xx') \rho(\xx')\notag\\
&\quad +\int d{\bf x} \rho({\bf x})U(\vct{x}).
\end{align}
The reader should note that ${\cal E}$ is not the free
energy functional of DFT \cite{bob_advances}.
Eq.~\eqref{dk1} may now be rewritten for identification of the stress
tensor $\boldsymbol \sigma$: The divergence of the stress tensor
appears directly \cite{Kreuzer,AerovKrugerJCP2014},
\begin{multline}
\frac{\partial\rho}{\partial t} ({\bf x},t) 
= -\beta D \nabla \cdot [\nabla \cdot \ssigma(\xx,t) - \rho(\xx,t) \nabla U(\xx)] \\
+\nabla\cdot \left[\sqrt{2D\rho(\xx,t)}{\boldsymbol \eta}({\bf x},t)\right].\label{dk2}
\end{multline}
Eq.~\eqref{dk2} is a force balance between external and inter-particle forces, where the latter are
expressed via the stress tensor. The divergence of the stress tensor
is thus identified by comparing Eqs.~\eqref{dk2} and \eqref{dk1}. Using Eq.~\eqref{eq:pot}, we obtain
\begin{align}\label{eq:ds}
\nabla\cdot\boldsymbol{\sigma}(\vct{x})=-k_BT\nabla \rho(\vct{x})- \rho(\vct{x})\nabla\int d\xx' V({\bf x}-\xx') \rho(\xx').
\end{align}
The divergence of the stress tensor at position $\vct{x}$ thus has a
local entropic or osmotic contribution involving the density at $\vct{x}$, and an
interaction term which involves the potential $V$. We emphasize that
Eq.~\eqref{eq:ds} gives the instantaneous stress tensor, which is
valid for any given (snapshot) configuration of particles. 

The noise-averaged form of Eq.~\eqref{eq:ds} can be rewritten using the two-body
density of Eq.~\eqref{eq:rho2}, which can also be expressed via
\begin{align}\label{eq:rho21}
	\left\langle \rho(\vct{x})\rho(\vct{x}')\right\rangle=\langle\rho(\vct{x})\rangle\delta(\vct{x}-\vct{x}')+\rho^{(2)}(\vct{x},\vct{x}').
\end{align}
Noting that the first term on the rhs of Eq.~\eqref{eq:rho21} does not
contribute in Eq.~\eqref{eq:ds} (reflecting the fact that a particle
cannot exert a force on itself), we obtain the familiar form for the
divergence of $\boldsymbol \sigma$ \cite{Kreuzer}, 
\begin{align}\label{eq:dsm}
	\langle\nabla\cdot\boldsymbol{\sigma}(\vct{x})\rangle=&-k_BT\nabla \langle\rho(\vct{x})\rangle\notag\\&- \int d{\bf x}' [\nabla V({\bf x}-{\bf x}')] \rho^{(2)}(\vct{x},{\bf x}').
\end{align}
This expression may for example be found starting from the Smoluchowski equation \cite{AerovKrugerJCP2014}.
The expression for the stress tensor itself, both instantaneous and
averaged, can be obtained from Eqs.~(\ref{eq:ds})
and~(\ref{eq:dsm}), respectively. Indeed, we show in Appendix~\ref{app:pr} that for
a spherical potential $V(\vct{r})=V(r)$ with $r=|\vct{r}|$, the
following expression of $\boldsymbol \sigma$ leads to the correct
force balance,
\begin{multline}
{\bf {\boldsymbol \sigma}}(\vct{x})=-k_BT\rho(\vct{x}) {\bf I}+\frac{1}{2}\int_0^1 d\lambda\int d\vct{r} \\
\times \frac{\vct{r}\vct{r}}{r} V'(r)\rho(\vct{x}+(1-\lambda)\vct{r})\rho(\vct{x}-\lambda\vct{r}) \:.\label{eq:sigmai}
\end{multline}
Averaging Eq.~\eqref{eq:sigmai}, we find for the mean stress tensor
\begin{multline}
\langle{\bf {\boldsymbol \sigma}}(\vct{x})\rangle=-k_BT\langle\rho(\vct{x})\rangle {\bf I}
+\frac{1}{2}\int_0^1 d\lambda\int d\vct{r} \\
\times \frac{\vct{r}\vct{r}}{r} V'(r)\rho^{(2)}(\vct{x}+(1-\lambda)\vct{r},\vct{x}-\lambda\vct{r}) \:,\label{eq:sigmaia}
\end{multline}
which is the celebrated Irving-Kirkwood formula~\cite{irving_kirkwood}
for the stress tensor; Eq.~(\ref{eq:sigmai}) extends it to
individual microscopic configurations. 
Note that adding a
divergence-free term to the stress tensor does not change the force
balance in Eq.~(\ref{dk2}) so that different expressions of the stress
tensor are acceptable. However, the expression of
Eq.~(\ref{eq:sigmaia}) can be argued to possess the most physical
symmetries~\cite{wajnryb_uniqueness_1995}.

\subsection{Stress tensor for the field $\phi$}
For further use, we also give the form of the stress tensor in
terms of the fluctuating field $\phi$, such that
$\rho=\bar\rho+\phi$. From Eq.~(\ref{eq:sigmai}), this directly gives
\begin{multline}
{\bf {\boldsymbol \sigma}}(\vct{x})=-k_BT\left[\bar\rho(\xx)+\phi(\vct{x})\right] {\bf I}+\frac{1}{2}\int_0^1 d\lambda\int d\vct{r} \times \\
\frac{\vct{r}\vct{r}}{r} V'(r)\left[\bar\rho+\phi\right](\vct{x}+(1-\lambda)\vct{r}) \left[\bar\rho+\phi\right](\vct{x}-\lambda\vct{r}) \:.\label{eq:sigmai3}
\end{multline}
If the correlations of the field $\phi$ are known (e.g. assuming 
Gaussian fluctuations~\cite{Kruger17}), Eq.~\eqref{eq:sigmai3} can
then be used to compute the correlations of the stress tensor, and the
viscosity via Green-Kubo relations.

Eq.~(\ref{eq:sigmai3}) can be simplified further to a form which displays clearly the off-diagonal
components, and which will be useful for computing shear viscosity. Up to
a divergence-free term, one gets for bulk systems, where $\bar\rho$ is not a function of $\vct{x}$, that (this may be directly inferred from Eq.~\eqref{eq:ds}) 
\begin{multline}
{\bf {\boldsymbol \sigma}}(\vct{x})=-\left(k_BT\phi(\xx)
+\bar\rho \int d\xx'\ V(\vct{x}-\xx')\phi(\xx')\right)
{\bf I} \\
  +\frac{1}{2}\int_0^1 d\lambda\int d\vct{r} 
\frac{\vct{r}\vct{r}}{r} V'(r)\phi(\vct{x}+(1-\lambda)\vct{r})\phi(\vct{x}-\lambda\vct{r})
\:.\label{eq:sigmaiphi}
\end{multline}
The above shows that only the terms quadratic in $\phi$ contribute to
the off-diagonal components of the stress tensor in bulk (and thus, as we will
see later, to the viscosity). While Eq.~\eqref{eq:sigmaiphi} is useful for computation of off-diagonal elements, care should be taken for the diagonal term in this representation, because the integral over the potential may become arbitrarily large (Eq.~\eqref{eq:sigmaiphi} omits diagonal terms due to $\bar\rho$ which are necessary to regularize the pair correlation for large $V$).

\subsection{The different terms in the force balance}
\label{sec:sigmaC}
The terms in Eq.~\eqref{dk2} are interpreted physically as force
densities. There is the force (density) acting on the external
potential $U$ (for example, the force acting on a wall which bounds the
fluid),
\begin{align}\label{eq:fU}
\vct{f}^{(U)}(\xx)=\rho(\xx)\nabla U(\xx).
\end{align}
In equilibrium, there is no net particle current in the system, so
that the external force balances on average the divergence of
$\boldsymbol{\sigma}$
\begin{align}\label{eq:sf}
\langle \vct{f}^{(U)}(\xx)\rangle^{\rm eq}=\left\langle\nabla\cdot\boldsymbol{\sigma}(\xx)\right\rangle^{\rm eq}.
\end{align}
Eq.~\eqref{eq:sf} reflects the well-known fact that, in equilibrium,
the stress tensor is directly related to the force acting on walls or
embedded objects. Out of equilibrium, the mismatch of $\vct{f}^{(U)}$
and $\nabla\cdot\boldsymbol{\sigma}$ gives rise to particle
currents. 

Taking the time derivative of the density operator in
Eq.~\eqref{eq:do}, one obtains
\begin{align}\label{eq:r2}
\frac{\partial\rho}{\partial t} (\vct{x},t)= -\nabla\cdot \left[\sum_\mu \vct{v}_\mu(t) \delta(\vct{x}-\vct{x}_\mu(t))\right],
\end{align}
where $\vv_\mu=d\xx_\mu/dt$.
The instantaneous current is thus identified to be 
\begin{align}\label{eq:j}
\vct{j}(\vct{x},t)= \sum_\mu \vct{v}_\mu(t) \delta(\vct{x}-\vct{x}_\mu(t)).
\end{align}
Comparing Eqs.~\eqref{eq:r2} and \eqref{dk2}, we obtain for the
mean current
\begin{align}\label{eq:fj}
\frac{k_BT}{D} \langle\vct{j}\rangle \equiv \langle\vct{f}^{(\vct{j})}\rangle=\langle\nabla\cdot\boldsymbol{\sigma}\rangle-\langle\vct{f}^{(U)}\rangle.
\end{align}
Eq.~\eqref{eq:fj} shows that 
$\nabla\cdot\boldsymbol{\sigma}$ and $\vct{f}^{(U)}$ balance the
frictional force $\vct{f}^{(\vct{j})}$ of that current~\cite{dhont}. Eq.~\eqref{eq:fj} has
yet another important consequence: Out of equilibrium, the stress
tensor $\ssigma$ may not generally be used to compute forces on walls
or embedded objects, because one needs to account for the contribution
of the currents.

\section{Stress tensor in Landau-Ginzburg theory}
\label{sec:sigmalandau}
In this section, we take a detour to consider the stress tensor for
field theories obeying the Landau-Ginzburg Hamiltonian. Indeed, the
form of the stress tensor in this context has been a subject of recent
debate~\cite{deangopinathan2009JStatMech,deangopinathan2010PRE,bitbolfournier2011forces}, and
insight can be gained by comparing with the microscopic stress
tensor of Eq.~\eqref{eq:sigmai} as well as to the discussion in
Sec.~\ref{sec:sigmaC}.

Phenomenological field theories are particularly well suited to
investigate large-scale generic phenomena. One may, for example, study
the universal aspects of systems near critical points
\cite{hohenberg,onukibook,kardarbook} or the properties of long-ranged
correlations which are present due to out-of-equilibrium initial
conditions as, e.g., in Ref.~\cite{Rohwer17}. In such scenarios, one
seeks expressions independent of the microscopic details (such as the
interaction potential $V$). We thus investigate in this section the
possibility of expressing the stress tensor directly at the
coarse-grained level of the field theory, based purely on the Landau-Ginzburg Hamiltonian. Let us consider the
Hamiltonian for a scalar field $\Phi$ in $d$ dimensions,
\begin{align}\label{eq:HLG}
{H}[\Phi]=\int d\xx\left[\frac{\kappa}{2}(\nabla \Phi)^2 + {\cal U}(\vct{x}) \right]\equiv\int d\xx \cal{H}(\vct{x}).
\end{align}
Although higher orders in $\nabla \Phi$ can be included based on symmetry arguments~\cite{kardarbook}, we do not consider this case here. $\mathcal U$ can be a general polynomial of $\Phi$, but the simplest example of the above field theory is the Gaussian case
where ${\cal U}(\vct{x}) = m(\vct{x})\Phi(\xx)^2/2$. Here, $m(\vct{x})$ can
be a function of position so that it may include contributions from
external potentials. For bulk, with $m$ spatially homogeneous, the correlation length  is then set by $\sqrt{\kappa/
  m}$~\cite{kardarbook}.

At the level of effective field theories such as Eq.~\eqref{eq:HLG},
the definition of mechanical quantities is not obvious. Indeed, one needs to define the nature of the field $\Phi$ and specify whether it corresponds to a matter field (for instance a particle density, a spin density or local charge density) or a potential field (for instance the local electrostatic potential or chemical potential). If the field $\Phi$ is in an arbitrary non-equilibrium configuration, in the mechanical sense, then there are  locally unbalanced body forces in the system. We proceed by applying a fictitious external field so that the system is in local mechanical equilibrium. The force exerted by this fictitious field on a volume therefore cancels out exactly the local body forces generated by the internal interactions in the system. Concretely, we apply an external field $h$ which shifts the overall energy (Hamiltonian) of $\Phi$ to
\begin{equation}
{H}_h[\Phi] = {H}[\Phi] + \int d\xx \ h({\bf x})\Phi({\bf x}).
\end{equation}
Next, for a given configuration of $\Phi$, the external field is
chosen to ensure local mechanical equilibrium  of the given configuration when the field is applied, {\it i.e.}
${\delta { H_h}\over \delta \Phi({\bf x}) }=0$. The required
field $h$ is thus
\begin{equation}\label{eq:h}
  h({\bf x}) = - {\delta { H}[\Phi]\over \delta \Phi({\bf x})}
\end{equation}
from which we can extract the force density ${\bf f}_h$ acting on the
subvolume $V$ due to the imposed field, which follows from its
fundamental definition from the Hamiltonian $H_h$, 
(see e.g. Ref.~\cite{deangopinathan2010PRE})
\begin{equation}\label{eq:field0}
{\bf f}_h(\xx) = - \Phi({\bf x}) \nabla h({\bf x}) =   \Phi({\bf x}) \nabla  {{\delta H[\Phi]}\over {\delta \Phi({\bf x})}}.
\end{equation}
If $\Phi$ is a density field, the force density may be viewed intuitively as the product of the density and the gradient of the chemical potential associated with $H[\Phi]$.

Because the force due to $h$ balances the force density that is not due to $h$, we obtain for the (body) force density $\vct{f}^{(\vct{j})}$ when $h=0$:
\begin{equation}\label{eq:field}
{\bf f}^{(\vct{j})}(\xx) = - \Phi({\bf x}) \nabla  {{\delta H[\Phi]}\over {\delta \Phi({\bf x})}}.
\end{equation} 
We added the superscript $\vct{j}$ in order to emphasize the similar
nature of the forces in Eq.~\eqref{eq:field} and Eq.~\eqref{eq:fj},
which will become more apparent below. 

We note that Eq.~\eqref{eq:field} is exactly the  force density in Eq.~\eqref{dk1}, when one makes the identification $\Phi\equiv\rho$ and $H=\cal{E}$. It is important  to note that if the field theory is written down in terms of the fluctuations of the field $\Phi$ about its average value $\overline \Phi$, as $\Phi = \overline \Phi+ \phi$, and the Hamiltonian for the fluctuations is $H_f[\phi]$, the local body force is given by
\begin{equation}
{\bf f}^{(\vct{j})}(\xx) = - \left[\overline \Phi(\xx)+\phi({\bf x})\right]\nabla  {{\delta H_f}\over {\delta \phi({\bf x})}}.
\end{equation} 
In the limit of small fluctuations the above can be used to derive an approximate model B dynamics for interacting particle systems \cite{deanpodgornik2014,lu15,dean16,demery16}, where
only linear terms in $\phi$ are kept, yielding
\begin{equation}
{\bf f}^{(\vct{j})}(\xx) 
= - \overline \Phi(\xx)\nabla\left[\int d\xx' \left.{{\delta^2 H_f}\over {\delta \phi({\bf x})\delta \phi(\xx')}}\right|_{\phi=0}\bf\phi(\xx')\right].
\end{equation}
Next, we identify the force on
the external potential, as in Eq.~\eqref{eq:fU}. Here, the force
density ${ \vct{f}}^{(U)}$ is identified by the change of ${\cal H}$
under a small displacement of the potential or, equivalently, a small
displacement of the object giving rise to the potential $U$
\cite{deangopinathan2010PRE}. Defining the vector $\vct{X}$ from the
origin to a randomly chosen point on the object, we have
\begin{align}\label{eq:forceO}
{ \vct{f}}^{(U)} = -\nabla_X {\cal H}.
\end{align}
Manipulating the partial and functional derivatives as detailed in
Appendix~\ref{app:GH}, we obtain the relation between the force
density in Eq.~\eqref{eq:field} and Eq.~\eqref{eq:forceO},
\begin{align}\label{eq:force}
 \vct{f}^{(\vct{j})} = \nabla\cdot \vct{T}- {\vct{f}}^{(U)},
\end{align}
where we have introduced the stress tensor $\vct{T}$,
\begin{equation}\label{eq:T}
T_{ij}(\xx) = \delta_{ij}\left( {\cal H}(\xx)-\Phi({\bf x}){\delta  {H}\over \delta\Phi({\bf x})}\right) -  {\partial {\cal H}\over \partial \nabla_j \Phi(\xx)} \nabla_i\Phi(\xx).
\end{equation}
(In this section we use $\bf T$ instead of $\boldsymbol \sigma$, following the usual
field theory notation.) The structure of Eq.~\eqref{eq:force} is identical to that in
Eq.~\eqref{eq:fj}, so that the field theory stress tensor from
Eq.~\eqref{eq:T} is on the same footing as the microscopic one. As mentioned below Eq.~\eqref{eq:fj}, the computation of forces on external objects, ${\vct{f}}^{(U)}$,
is non-trivial out of equilibrium because of the term involving the
current in Eq.~\eqref{eq:force}. This will be relevant, for example,
when computing Casimir forces between walls or objects. As a
crosscheck, we show in Appendix \ref{app:GH} that Eq.~\eqref{eq:force}
gives the same value for ${\vct{f}}^{(U)}$ as the expression
previously derived in Eq.~(18) of Ref.~\cite{deangopinathan2010PRE}.  A particular case worth mentioning is that of no-flux boundary conditions, where the surface normal component of $\vct{f}^{(\vct{j})}$ is forced to vanish. For the geometry of two parallel plates, the force (or pressure) acting on the plates can then be computed by evaluating the stress tensor in Eq.~(\ref{eq:T}) at the given surface. This was used for computation of non-equilibrium Casimir forces in Ref.~\cite{Rohwer17} and tested quantitatively with simulations of interacting Brownian particles in Ref.~\cite{Rohwer17c}. 

In global equilibrium, one can show~\cite{deangopinathan2010PRE} that
$\vct{f}^{(\vct{j})}$ vanishes on average (see Eq.~\eqref{eq:veq}), as
was the case in the microscopic theory in Eq.~\eqref{eq:fj}. Our
formulation (and the form of $\bf T$) then agrees with the commonly
accepted equilibrium definition of the stress tensor $\vct{T}^{\rm eq}$, i.e.,
\begin{align}\label{eq:forceeq}
\langle\nabla\cdot \vct{T}\rangle^{\rm eq}= \langle\nabla\cdot \vct{T}^{\rm eq}\rangle^{\rm eq}= \langle{\vct{f}}^{(U)}\rangle^{\rm eq}
\end{align}
and $\vct{T}^{\rm eq}$ expressed as (using Eq.~\eqref{eq:veq})
\begin{equation}\label{eq:Teq}
T^{\rm eq}_{ij}= \delta_{ij} {\cal H} -  {\partial {\cal H}\over \partial \nabla_j \Phi} \nabla_i\Phi.
\end{equation}

\section{Application: Stress correlations of Brownian suspensions}\label{sec:App}
In this section, we make use of the expressions derived for the
microscopic stress tensor Eq.~\eqref{eq:sigmai3} to compute the
two-point correlations of the stress at different positions and times
for a bulk equilibrium system. We provide the limit of high
temperatures, which corresponds to the leading-order term in the external potential $V$.
We also compare with the results obtained from the Gaussian field theory,
using Eq.~\eqref{eq:T} to define the stress tensor.

\subsection{Diagonal Components -- pressure fluctuations}
We start with the diagonal part of the stress tensor in
Eq.~\eqref{eq:sigmai3}, denoting $\delta {\boldsymbol
  \sigma}={\boldsymbol\sigma}- \langle{\boldsymbol\sigma}\rangle^{\rm
  eq}$. The leading term at small $V$ (or high $T$) results from the
ideal contribution in Eq.~\eqref{eq:sigmai3}. It reads (without
summing repeated indices),
\begin{align}
\lim_{\beta V\to0}\left\langle\delta\sigma_{ii}(\vct{x},t) \delta\sigma_{kk}(\vct{0},0) \right\rangle^{\rm eq}= (k_BT)^2  \left\langle  \phi(\vct{x},t) \phi(\vct{0},0)\right\rangle^{\rm eq}. \label{eq:stp}
\end{align}
In order to determine the leading order in $V$, the correlation function in Eq.~\eqref{eq:stp} should be evaluated for $V=0$, which corresponds to the ideal gas. 
The correlations for the ideal gas are denoted $\langle \cdot \rangle^\mathrm{id}$ and are computed in Appendix~\ref{app:id}.
The two-point correlation reads
\begin{align}\label{eq:C}
\left\langle  \phi(\vct{x},t) \phi(\vct{0},0)\right\rangle^{\rm id}= \frac{\bar\rho}{(4\pi  D t)^{3/2}} e^{-\frac{|\vct{x}|^2}{4  D t}}.
\end{align}
We obtain
\begin{align}\label{eq:dd}
\lim_{\beta V\to0}\left\langle\delta\sigma_{ii}(\vct{x},t) \delta\sigma_{kk}(\vct{0},0) \right\rangle^{\rm eq} = \frac{(k_BT)^2\bar\rho}{(4\pi  D t)^{3/2}} e^{-\frac{|\vct{x}|^2}{4  D t}} . 
\end{align}
We observe from Eq.~\eqref{eq:dd} that the temporal correlations are
due to a diffusion process. The long-ranged character of the
correlations in space becomes more apparent after integrating Eq.~(\ref{eq:dd}) over time
\begin{align}\label{eq:lr}
\int_0^\infty dt \lim_{\beta V\to0}\left\langle\delta\sigma_{ii}(\vct{x},t) \delta\sigma_{kk}(\vct{0},0) \right\rangle^{\rm eq} = \frac{(k_BT)^2 \bar\rho}{4\pi  D |\vct{x}| }. 
\end{align}
This quantity decays slowly, as the inverse of the distance. 
Because the time integral of such equilibrium correlations appears
typically in linear response formulae, we expect the long-ranged form of
Eq.~\eqref{eq:lr} to be relevant for non-equilibrium perturbations.

As a final note, the equal-time fluctuations of pressure from
Eq.~\eqref{eq:dd} are
\begin{align}
\int d\vct{x}\lim_{\beta V\to0}\left\langle\delta\sigma_{ii}(\vct{x},0) \delta\sigma_{kk}(\vct{0},0) \right\rangle^{\rm eq}=(k_BT)^2\bar\rho.
\end{align}
These are the pressure fluctuations of an ideal gas, which have a
partly controversial history, starting with Gibbs \cite{Gibbs02}.

We note that the stress tensor of Landau-Ginzburg theory,
Eq.~\eqref{eq:T}, together with a Gaussian Hamiltonian,
Eq.~\eqref{eq:HLG}, yields a qualitatively different result for
spatial pressure correlations. The reason is that Eq.~\eqref{eq:T} 
yields a stress tensor which is purely quadratic in
the fluctuating field $\Phi$, resulting from the symmetry assumptions underlying the field theoretic Hamiltonian in Eq.~\eqref{eq:HLG} \cite{kardarbook}. In contrast to that, it is the term $\sim\rho\ln\rho$ in Eq.~\eqref{eq:pot} which yields the correct ideal gas contribution.

\subsection{Diagonal--off-diagonal correlations}

As visible in Eq.~\eqref{eq:sigmaiphi}, the off-diagonal component of
the stress involves the potential $V$. The leading term of the
correlation between diagonal and
off-diagonal components of $\boldsymbol \sigma$ is thus linear in $V$. This term reads (for
$i\not=j$)
\begin{align}
&\lim_{\beta V\to0}\left\langle\delta\sigma_{ij}(\vct{x},t) \delta\sigma_{kk}(\vct{0},0) \right\rangle^{\rm eq}= k_BT \int_0^1  d\lambda\int d\vct{r} \notag\\
&\times \frac{r_ir_j}{r} V'(r)\left\langle \phi(\vct{\vct{x}+(1-\lambda)\vct{r}}),t) \phi(\vct{x}-\lambda\vct{r},t)\phi(\vct{0},0)\right\rangle^{\rm id}, \label{eq:stp2}
\end{align}
where, again, the correlation of $\phi$ is evaluated for the ideal gas.
Since the density of the ideal gas has Poissonian statistics~\cite{velenich08}, the three point correlator in Eq.~\eqref{eq:stp2} is per se non-zero.
Using Eq.~\eqref{eq:rho3}, we find the correlation in terms of Fourier transforms,
\begin{equation}
\langle \tilde \phi({\bf k},t)\tilde \phi({\bf k}',t)\tilde \phi({\bf q},0)\rangle^{\rm id} =  (2\pi)^3 \overline\rho\delta({\bf k}+{\bf k'}+{\bf q})e^{-Dq^2 t},\label{eq:phi3}
\end{equation}
where, in this paper, Fourier transforms are defined by
\begin{equation}
\tilde f(\vct{k})=\int d\vct{x} f(\vct{x})e^{-i{\bf k}\cdot{\bf x}}.
\end{equation}
While there is indeed a non-zero three point correlation in Eq.~\eqref{eq:phi3}, its contribution to the correlation of stresses in Eq.~\eqref{eq:stp2} is zero (Appendix~\ref{app:id}). 
The reason is that Eq.~\eqref{eq:phi3} results from a single Brownian particle and its self-correlations (as is apparent from the linearity in $\bar\rho$), and this particle cannot exert a force  upon itself. Some relations useful to derive this statement mathematically are given in Appendix \ref{app:id}. We thus note that there is no correlation between diagonal and off-diagonal stress components at the given order in $V$.
 
\subsection{Off diagonal components -- shear stress fluctuations}
Here we start by computing the exact leading term in $V$ in the
microscopic theory. We then also derive the shear stress fluctuations
in the Gaussian Landau-Ginzburg theory.

\subsubsection{Exact leading order in $V$}
To leading order, the off-diagonal component of the stress correlator is proportional to $V^2$. For ease of notation, we abbreviate,
\begin{equation}
\lim_{\beta V\to 0}\left\langle\delta\sigma_{ij}(\vct{x},t) \delta\sigma_{kl}(\vct{0},0) \right\rangle = \Sigma_{ijkl}(\vct{x},t).
\end{equation}
For $i\not=j$ and $k\not=l$, this expression reads
\begin{align}
\notag&\Sigma_{ijkl}(\vct{x},t)=\frac{1}{4}  \int d{\bf r}d{\bf r}' \int_0^1 d\lambda d\lambda'
\frac{r_i r_j}{r}V'(r) \frac{r'_k r'_l}{r'}V'(r')\times\\
&\left\langle \phi(\vct{x}+\vct{r}-\lambda\vct{r},t)¸\phi(\vct{x}-\lambda\vct{r},t) \phi(\vct{r}'-\lambda'\vct{r}',0)\phi(-\lambda'\vct{r}',0)\right\rangle^{\rm id}\!.
\label{eq:sigsig}
\end{align}
In Fourier space, the correlation reads (again, see Appendix~\ref{app:id}),
\begin{align}
&\langle \tilde \phi({\bf k},t) \tilde \phi({\bf k}',t) \tilde \phi({\bf q},0) \tilde \phi({\bf q}',0) \rangle^{\rm id} \nonumber\\
&\quad = \overline \rho (2\pi)^3\delta({\bf k} + {\bf k}' +{\bf q}+{\bf q}') {\cal C}({\bf k}+{\bf k}',t)\nonumber\\
&\qquad +(2\pi)^6\overline \rho^2[ \delta({\bf k}+{\bf k}')\delta({\bf q}+{\bf q}')\nonumber\\
&\qquad+ \delta({\bf k}+{\bf q})\delta({\bf k}'+{\bf q}'){\cal C}({\bf k},t){\cal C}({\bf k}',t)\nonumber\\
&\qquad+\delta({\bf k}+{\bf q}')\delta({\bf k}'+{\bf q}){\cal C}({\bf k},t){\cal C}({\bf k}',t) ]\label{eq:phi4},
\end{align}
with ${\cal C}(\vct{k},t)=e^{-Dk^2t}$. The four-point correlation function in Eq.~\eqref{eq:phi4} contains a term linear in $\bar\rho$, which, as in Eq.~\eqref{eq:phi3} shows the non-Gaussian behavior of ideal particles. Again, this term does not contribute to the stress correlator. The last three terms in Eq.~\eqref{eq:phi4} are equivalent to a Gaussian decoupling, where all pairs of functions $\phi$ appear as in Wick's theorem. The first of these three is time independent and does not contribute to  Eq.~\eqref{eq:sigsig} by symmetry. We then find for $i\not=j$ and $k\not=l$
\begin{multline}
\Sigma_{ijkl}(\vct{x},t)= \int \frac{d\vct{k}}{(2\pi)^3} e^{i\vct{k}\cdot\vct{x}} \int_0^1 d\lambda d\lambda'\times\\
\frac{\overline\rho^2}{2}
\int \frac{d{\bf p}}{(2\pi)^3}\tilde{\cal{A}}_{ij}(\lambda {\bf k}-{\bf p})\tilde {\cal A}_{kl}(-\lambda' {\bf k}+{\bf p}) e^{-D [{\bf p}^2+(\vct{k}-\vct{p})^2]t},\label{sm}
\end{multline}
where for $i\neq j$ 
\begin{equation}
\tilde {\cal A}_{ij}(\vct{k})= \frac{k_ik_j}{k} \tilde V'(k).
\end{equation}

We continue by analyzing Eq.~\eqref{sm}
 in real space, which is found easily from the Gaussian decoupling of the four-point function in
Eq.~\eqref{eq:sigsig}. We thus have, using Eq.~\eqref{eq:C} (again $i\not=j$),
\begin{align}
&\Sigma_{ijij}(\vct{x},t)=\frac{1}{2}  \int d{\bf r}d{\bf r}' \int_0^1 d\lambda d\lambda'
\frac{r_ir_j}{r}V'(r) \notag\frac{r'_i r'_j}{r'}V'(r')\times\\
&  \frac{\bar\rho^2}{(4\pi  D t)^{3}} e^{\frac{-|\vct{x+\vct{r}-\lambda\vct{r}-\vct{r}'+\lambda' \vct{r}'}|^2-|\vct{x-\lambda\vct{r}+\lambda' \vct{r}'}|^2}{4  D t}} . 
\label{eq:sigsig2}
\end{align}

We now address the large distance behavior of the stress correlations.
For a short-ranged potential, the values of $|\rr|$ and $|\rr'|$ that
contribute to the above integral are constrained to that range. For
values of $|\vct{x}|$ much larger than the interaction range, we may
expand the exponential in Eq.~\eqref{eq:sigsig2} for $|\rr|\ll|\vct{x}|$
and $|\rr'|\ll|\vct{x}|$.  Because of the symmetry of the integrands in
Eq.~\eqref{eq:sigsig2}, we seek the leading terms of order
$r_ir_jr'_ir'_j$ of the exponential in
Eq.~\eqref{eq:sigsig2}. The integrals over $\lambda$ and $\lambda'$
can then be performed to obtain
\begin{multline}
	\Sigma_{ijij}(\vct{x},t) = \mathcal{G}[V] \\\times \bar\rho^2\frac{10  D^2 t^2- D t (x_i^2+x_j^2)+x_i^2x_j^2}{9216( D t)^7\pi^3} e^{-\frac{|\vct{x}|^2}{2 D t}} ,
\end{multline}
where we have defined the functional
\begin{equation}
	\mathcal{G}[V] =\frac{8\pi^2}{225}\left(\int_0^\infty dr r^5 V'(r)\right)^2.
\end{equation}
Here, too, we perform the time integral, which is the important
quantity in Green-Kubo relations, and obtain
\begin{equation}\label{eq:shf}
	\int_0^\infty dt\Sigma_{ijij}(\vct{x},t) = \mathcal{G}[V] \bar\rho^2\frac{5+40\frac{x_i^2x_j^2}{|\vct{x}|^4}-4\frac{x_i^2+x_j^2}{|\vct{x}|^2}}{48 D \pi^3 |\vct{x}|^8}. 
\end{equation}
Shear stress fluctuations in Eq.~\eqref{eq:shf} and pressure
fluctuations in Eq.~\eqref{eq:lr} thus decay as power laws in
space. These observations should have implications on the rheology of
suspensions in confined systems \cite{Evans81,aerov2015,Rohwer17b}. 

The long-ranged property of the shear stress correlations has been noted
in Ref.~\cite{Maier17} via a Zwanzig-Mori approach. In particular, a Fourier inversion of Eq.~(6) from Ref.~\cite{Maier17} with $s=0$ yields the analogue of $\int_0^\infty dt \notag\Sigma_{ijij}(\vct{x},t)$, giving contributions $\sim|\vct{x}|^{-3}$. These show a different power than those in Eq.~(\ref{eq:shf}), $\sim|\vct{x}|^{-8}$, which could be due to the absence of momentum conservation in our formulation.

\subsubsection{From Landau-Ginzburg Theory}

Next we consider the off-diagonal stress correlations from Landau-Ginzburg theory, using the stress tensor $\vct{T}$ in Eq.~\eqref{eq:T}. This,
together with a Gaussian decoupling of the four point correlation (again,
$i\not=j$), gives
\begin{multline}
\label{eq:sigLG}
\left\langle T_{ij}(\vct{x},t) T_{ij}(\vct{0},0) \right\rangle\approx \kappa \biggl([\nabla_i\nabla_j C(\vct{x},t)]^2\\
+[\nabla_i\nabla_i C(\vct{x},t)][\nabla_j\nabla_j C(\vct{x},t)]\biggr).
\end{multline}
We are interested in the case were $|\bf x|$ is much larger than the correlation
length. In this limit, the correlation function reads in model $B$
dynamics \cite{kardarbook,Kruger17} (see Appendix \ref{sec:eH} for
the leading order in correlation length) 
\begin{align}\label{eq:Cm}
C(\vct{x},t)=\frac{k_BT}{m} \frac{1}{(4\pi \tilde D t)^{3/2}} e^{-\frac{|\vct{x}|^2}{4 \tilde D t}},
\end{align}
where $m$ is the coefficient appearing in the Gaussian Hamiltonian in
Eq.~\eqref{eq:HLG} of the field theory. Explicitly, $m$ may also be expressed in terms of the 
small wave-vector limit of the direct correlation function $c^{(2)}$
\cite{HansenMcDonald, Kruger17} (see Appendix \ref{sec:eH}),
\begin{align}\label{eq:m_0}
m=k_BT\left(\frac{1}{\bar\rho}- c^{(2)}_0\right),
\end{align}
which may also be related to the isothermal compressibility
\cite{HansenMcDonald}.  
We have also introduced $\tilde D\equiv D/S_0$, with $S_0 = (1-\bar\rho c^{(2)}_0)^{-1} =  k_B T/\bar\rho m$ being the small wave-vector limit of the structure factor; $\tilde D$ is an approximation for the
diffusion coefficient \cite{dhont}, which emerges when taking the small wave-vector limit of Eq.~\eqref{eq:eom3}. Eq.~\eqref{eq:sigLG} yields
\begin{multline}
\label{eq:sigsig3}
\big\langle T_{ij}(\vct{x},t)  T_{ij}(\vct{0},0) \big\rangle =\\
\kappa^2\frac{(k_BT)^2}{m^2}\frac{ 2 \tilde D^2 t^2- \tilde D t (x_i^2+x_j^2)+x_i^2x_j^2}{512( \tilde D t)^7\pi^3} e^{-\frac{|\vct{x}|^2}{2 \tilde D t}} .
\end{multline}
Apart from prefactors, this expression is qualitatively equal to
Eq.~\eqref{eq:sigsig2}: It can be shown that the dependencies on $\bf
x$ and $t$ of Eqs.~\eqref{eq:sigsig3} and \eqref{eq:sigsig2} differ
only by terms which arise from divergence-free parts of the stress
tensor. 

\subsection{Viscosity}
We finish with the evaluation of the shear viscosity $\eta$ to leading
order in the interaction potential. Using the Green-Kubo relation \cite{kubobook}, the viscosity is given in terms of the stress correlator ($i\not=j$), 
\begin{equation}
\eta = \frac{\beta}{\mathcal V}\int_0^\infty dt \int d{\bf x} d{\bf x}' \langle \sigma_{ij}({\bf x},t)\sigma_{ij}({\bf x}',0)\rangle^{\rm eq},\label{eq:GK}
\end{equation}
where $\mathcal V$ is the volume of the system. 
Performing one spatial integral, one obtains
\begin{equation}
\eta =\beta \int d{\bf x} \int_0^\infty dt\langle \sigma_{ij}({\bf x},t)\sigma_{ij}({\bf 0},0)\rangle^{\rm eq}.\label{eq:GK2}
\end{equation}
Importantly, one should not integrate the result of
Eq.~\eqref{eq:shf} over $\vct{x}$ to obtain the result of
Eq.~\eqref{eq:GK2}, because Eq.~\eqref{eq:shf} is only valid for large
$\vct{x}$.
Instead, the integration over $\xx$ is straightforward from Eq.~\eqref{sm}, since it selects the mode $\kk=0$ in the integral, leading to
\begin{align}
\eta &=\frac{\beta{\overline \rho}^2}{4D} \int \frac{d{\bf p}}{(2\pi)^3}\frac{p_i^2p_j^2}{p^4}\tilde V'(p)^2\notag\\& = \frac{\beta{\overline\rho}^2}{120 \pi^2 D} \int_0^\infty dp p^2\tilde V'(p)^2.\label{eq:eta}
\end{align}
This is the exact result for the viscosity of Brownian suspensions to
leading order in the potential $V$, which is naturally quadratic. This
expression is also in agreement with perturbation computations, where
the perturbed pair distribution is found under shear. While such results have been
obtained for hard spheres \cite{brady_morris}, we have not been able
to find the result of Eq.~\eqref{eq:eta} in the literature for a
direct check.

As pointed out, the stress tensor is not uniquely defined, but only up to divergence-free terms. The viscosity in Eq.~\eqref{eq:eta} is, however, a measurable material property, which should be computable unambiguously. We therefore demonstrate that divergence-free terms do not contribute to Eq.~\eqref{eq:GK2}. First we note that for any vector (or tensor) field $\vct{A}(\vct{x})$ which decays quicker than $1/|\vct{x}|^3$ for large $|\vct{x}|$, we have 
\begin{align}
\int d{\bf x}  A_i(\vct{x})=   \int d{\bf x} (\nabla_j x_i) A_j(\vct{x})= -\int d{\bf x}  x_i \nabla_j A_j(\vct{x}).
\end{align}
We have used that $\nabla_j x_i=\delta_{ij}$. In the last step, partial integration was performed using the  decay properties of  $\vct{A}(\vct{x})$ at large $|\vct{x}|$, so that boundary terms vanish.
This  yields the relation
\begin{align} \label{eq:id}
\int d{\bf x}  \vct{A}(\vct{x})=-\int d{\bf x} \vct{x} \nabla\cdot\vct{A}. 
\end{align}
Proceeding by applying this to the  auto-correlator of stress in Eq.~\eqref{eq:GK2}, we denote $\vct{A}(\vct{x})=\int_0^\infty dt\langle \delta{\boldsymbol \sigma}({\bf x},t)\delta\sigma_{ij}({\bf 0},0)\rangle^{\rm eq}$. If this expression decays quickly enough, any divergence-free parts do not contribute to the viscosity in Eq.~\eqref{eq:GK2}. Using the form of $\vct \ssigma$ from Eq.~\eqref{eq:sigmaiphi} with $i\not=j$, we see that, to the considered order, the stress correlation decays quickly enough, and that therefore the viscosity in Eq.~\eqref{eq:eta} is unambiguous.
\section{Summary}
The exact stochastic equation for the evolution of  of the density operator \cite{Dean96} allows one to identify the (stochastic) form of the stress tensor directly. The mean of this quantity naturally agrees with the one found from the Smoluchowski
equation. It is interesting to note that non-equilibrium states with
a finite particle current give rise to an additional force
contribution, so that the stress tensor has to be applied with care
when computing forces on boundaries or objects. The non-equilibrium body force in a general field theory, where the field $\Phi$ is associated with a density operator, can be derived using force balance arguments. The result of this agrees with that obtained from the evolution of the 
density operator for the field theory given by Eq. (\ref{eq:pot}) with the identification $\Phi=\rho$. This allows a systematic derivation of the field theoretic stress tensor, valid out of equilibrium, and the associated procedure for computing non-equilibrium forces on immersed objects/potentials.

We further computed the spatial and temporal correlations of the stress tensor exactly to leading order in interaction potential $V$. These
correlations (or more generally forces) are more long-ranged than the
equilibrium correlation length, a fact which we attribute to the
conservation of particles in the system considered here. In contrast to
Ref.~\cite{Maier17}, our system conserves particle density, but not
momentum. Off-diagonal stress correlations from the field theory were shown to agree qualitatively with those from the microscopic stress tensor.

Future work can thus involve computing these correlations for the case of
particle density and momentum conservation. The impact of such correlations on flow in confinement poses further interesting questions \cite{aerov2015,Rohwer17b}.

\begin{acknowledgments}
  This research was supported by the DFG grant No.~KR 3844/2-1 and the
  ANR project FISICS. A.S.~acknowledges funding through a PLS
  fellowship of the Gordon and Betty Moore foundation. C.M.R.~gratefully acknowledges S.~Dietrich for financial support. V.D. thanks M.~Schindler for interesting
discussions.
\end{acknowledgments}

\begin{appendix}
\section{Derivation of the stress tensor for any realization of the density field  (Eq.~\eqref{eq:sigmai})}\label{app:pr}
Here we show  that the divergence of Eq.~\eqref{eq:sigmai} equals Eq.~\eqref{eq:ds} under minimal assumptions and without any thermodynamic averaging. An alternative derivation can be found in Ref.~\cite{percuspaper}. Making the trivial decomposition into an ideal part and the interacting term, we can write
\begin{equation}
\boldsymbol{\sigma} = \boldsymbol{\sigma}^i+\boldsymbol{\sigma}^{ni}
\end{equation}
where $\boldsymbol{\sigma}^i$ is the ideal gas contribution:
\begin{equation}
\boldsymbol{\sigma}^i({\bf x}) = -k_BT \rho({\bf x}){\bf I}.
\end{equation}
The remaining non-ideal term due to inter-particle interactions can be written as 
\begin{multline}
\nabla\cdot \boldsymbol{\sigma}^{ni}(\xx) = -\rho({\bf x}) \int d{\bf r} \nabla V({\bf r})\rho({\bf x}-{\bf r})\\
=-\frac{1}{2}\rho({\bf x}) \int d{\bf r} \left[\nabla V({\bf r})\rho({\bf x}-{\bf r})+\nabla V(-{\bf r})\rho({\bf x}+{\bf r})\right],
\end{multline}
where we used a translation of the integration variable, and in the second line we have symmetrized the integral over the variables ${\bf r}$ and $-{\bf r}$. At this point it is important to choose interaction potentials obeying the
reflection symmetry $\nabla V({\bf r})= -\nabla V(-{\bf r})$ (this is clearly satisfied by isotropic potentials with $V(\vct{r})=V(r)$), for which one may write
\begin{align}
\nabla V(\vct{r})=\frac{\vct{r}}{r} V'(r).
\end{align}
This leads to
\begin{align}
&\nabla\cdot \boldsymbol{\sigma}^{ni}(\xx) =-\frac{1}{2} \int d{\bf r} \nabla V({\bf r})\left[\rho({\bf x})\rho({\bf x}-{\bf r})-\rho({\bf x})\rho({\bf x}+{\bf r})\right].
\end{align}
Denoting 
\begin{equation}
f({\bf x},{\bf r},\lambda) = \rho(\vct{x}-\lambda\vct{r})\rho(\vct{x}+(1-\lambda)\vct{r}),
\end{equation}
we can write
\begin{align}
\nabla\cdot \boldsymbol{\sigma}^{ni}(\xx) & =-\frac{1}{2} \int d{\bf r} \nabla V({\bf r})\left[
f({\bf x},{\bf r},1)-f({\bf x},{\bf r},0)\right]\\
&=-\frac{1}{2} \int d{\bf r} \nabla V({\bf r})\left[
\int_0^1d\lambda \frac{\partial f({\bf x},{\bf r},\lambda)}{\partial \lambda}\right].
\end{align}
It is easy to see that
\begin{equation}
\frac{\partial f({\bf x},{\bf r},\lambda)}{\partial \lambda}= -\vct{r}\cdot\nabla_{\bf x} \left[\rho(\vct{x}-\lambda\vct{r})\rho(\vct{x}+(1-\lambda)\vct{r})\right],
\end{equation}
where the gradient should be taken with respect to ${\bf x}$. Putting this together for isotropic potentials then yields
\begin{align}
&\nabla\cdot \boldsymbol{\sigma}^{ni}(\xx)\notag\\
&=\nabla \cdot \left[ \frac{1}{2} \int d{\bf r} \frac{\vct{r}\vct{r}}{r} V'(r)\int_0^1d\lambda \rho(\vct{x}-\lambda\vct{r})\rho(\vct{x}+(1-\lambda)\vct{r})\right].
\end{align}
The equality of the divergence of Eqs.~\eqref{eq:sigmai} and~\eqref{eq:ds} then follows.

\section{Relations used for the derivation of the stress tensor in the field theory}\label{app:GH}
The following relation is useful for derivation of Eq.~\eqref{eq:force}  (noting that the Hamiltonian depends on the field and its gradients), 
\begin{equation}
\nabla_i {\cal H} = (\nabla_i\Phi) {\partial {\cal H}\over \partial \Phi} + (\nabla_i\nabla_j\Phi) {\partial {\cal H}\over \partial\nabla_j\Phi}-\frac{\partial {\cal H}}{\partial X_i}.
\end{equation}
We have added a term in the Hamiltonian, which depends explicitly on a position ${\bf X}$,  the position of the object giving rise to the external potential $U$.
 Another explicit form is 
 \begin{equation}
{\delta { H}\over \delta \Phi({\bf x})} = {\partial {\cal H}\over \partial \Phi}-{\nabla}_j{\partial {\cal H}\over \partial \nabla_j\Phi}.
\end{equation}
Putting this together yields for the force in Eq.~\eqref{eq:field}
\begin{align}
\notag f_i^{(\vct{j})} =    - \nabla_i \left[\Phi({\bf x}) {\delta { H}\over \delta \Phi({\bf x})}\right] + \nabla_i {\cal H} +\frac{\partial {\cal H}}{\partial X_i}\\
-\nabla_j\left[ (\nabla_i\Phi) {\partial {\cal H}\over \partial \nabla_j\Phi}\right]. 
\end{align}
We can also identify $-\frac{\partial {\cal H}}{\partial X_i}$ with the force density in direction $i$ acting on the object described by the coordinate $\bf X$.
This gives Eqs.~\eqref{eq:force} and \eqref{eq:T} in the main text. 

In order to demonstrate the agreement between Eq.~\eqref{eq:force} and Ref.~\cite{deangopinathan2010PRE},  we  use
\begin{align}\label{eq:iden}
 \nabla_i \left(\Phi({\bf x}) {\delta {\mathcal H}\over \delta \Phi({\bf x})}\right)=\Phi({\bf x}) \nabla_i {\delta {\mathcal H}\over \delta \Phi({\bf x})}+{\delta {\mathcal H}\over \delta \Phi({\bf x})}
\nabla_i\Phi({\bf x})
\end{align}
to arrive at 
\begin{equation}\label{eq:FO}
 f^{(U)}_{i} =  \nabla_j \left[\delta_{ij}{\cal H}- \nabla_i\Phi {\partial {\cal H}\over \partial \nabla_j \Phi}\right] -{\delta {H}\over \delta \Phi({\bf x})}
\nabla_i\Phi({\bf x}),
\end{equation}
which is identical to Eq.~(18) of Ref.~\cite{deangopinathan2010PRE}.

In global equilibrium, one can prove \cite{deangopinathan2010PRE} that the body force vanishes exactly, and, furthermore, that 
\begin{align}\label{eq:veq}
\nabla_i\left\langle \Phi({\bf x}) {\delta {\mathcal H}\over \delta \Phi({\bf x})}\right\rangle^{\rm eq}=0= \left\langle\Phi({\bf x})\nabla_i {\delta {\mathcal H}\over \delta \Phi({\bf x})}\right\rangle^{\rm eq}.
\end{align}
\section{Correlations of an ideal gas}\label{app:id}
For the computation of the stress tensor correlations to leading order in potential $V$, we require the time dependent correlation functions of the ideal gas, which can be computed exactly \cite{velenich08}. 
We write the Fourier transform of the density field for $N$ ideal particles as
\begin{equation}
\tilde \rho({\bf k},t) = \sum_{\mu=1}^N \exp\left(-i{\bf k}\cdot[{\bf x}_{\mu0} + {\bf X}_\mu(t)]\right),
\end{equation}
where ${\bf x}_{\mu0}$ denotes the position of particle $\mu$ at time $t=0$ and ${\bf X}_\mu(t)$ is its subsequent displacement at time $t$. In this approximation all the ${\bf X}_\mu(t)$ are independent Brownian motions with diffusion constant $D$. Two averages are now taken. Firstly, the average over the initial coordinate ${\bf x}_{i0}$ takes the form
\begin{equation}
\langle \cdot\rangle = \frac{1}{\mathcal V}\int d{\bf x} \cdot,
\end{equation}
where $\mathcal V$ is the volume of the system. The second average is over the Brownian motions ${\bf X}_\mu(t)$. The first two correlation functions are
\begin{equation}
\langle \tilde \rho({\bf k},t)\rangle^{\rm id} = \overline \rho (2\pi)^3 \delta({\bf k})
\end{equation}
and
\begin{multline}
\langle \tilde \rho({\bf k},t)\tilde \rho({\bf q},0)\rangle^{\rm id}\\
= (2\pi)^3 \overline \rho\delta({\bf k}+{\bf q}) e^{-D q^2 t}
+(2\pi)^6 \overline{\rho}^2\delta({\bf k})\delta({\bf q}).\label{eq:2p}
\end{multline}
From these correlations, we can get the correlations of the fluctuations.
First, we note that the decomposition $\rho(\xx,t)=\bar\rho+\phi(\xx,t)$ reads in Fourier space $\tilde\rho(\kk,t)=(2\pi)^3\bar\rho\delta(\kk)+\tilde \phi(\kk,t)$.
We deduce that
\begin{align}
\langle \tilde\phi(\kk,t) \rangle^{\rm id} & = 0,\\
\langle \tilde\phi(\kk,t)\tilde\phi(\qq,0) \rangle^{\rm id} & = (2\pi)^3 \overline \rho\delta({\bf k}+{\bf q}) e^{-D q^2 t}. \label{eq:2p_phi}
\end{align}
Eq.~\eqref{eq:2p_phi} is used to get Eq.~\eqref{eq:C} from Eq.~\eqref{eq:stp} in the main text.

The three point function is found similarly. After a few computation steps, we get   
\begin{widetext}
\begin{multline}\label{eq:rho3}
\langle \tilde \rho({\bf k},t)\tilde \rho({\bf k}',t)\tilde \rho({\bf q},0)\rangle^{\rm id}
=  (2\pi)^3 \overline\rho\delta({\bf k}+{\bf k'}+{\bf q})\exp(-Dq^2 t)\\
+ (2\pi)^6 \overline\rho^2 \left[\delta({\bf k}+{\bf k'})\delta({\bf q})
+ \delta({\bf k}+{\bf q})\delta({\bf k}')
+ \delta({\bf k}'+{\bf q})\delta({\bf k})\right] \exp(-Dq^2 t)
+ (2\pi)^9 \overline\rho^3\delta({\bf k})\delta({\bf k}')\delta({\bf q}).
\end{multline}
\end{widetext}
The three-point correlation of the fluctuations reduce to Eq.~\eqref{eq:phi3} in the main text. 

A useful relation is 
\begin{equation}
\tilde A_{ij}({\bf k})= -\delta_{ij}\tilde V(k) -\frac{k_i k_j}{k}\tilde V'(k)
\end{equation}
with inverse Fourier transform
\begin{equation}
A_{ij}({\bf r}) = \frac{r_ir_j}{r}V'(r).\label{defa}
\end{equation}
We also note that $\tilde A_{ij}({\bf k}) = \tilde A_{ij}(- {\bf k})$. The following relation reflects the fact that a particle cannot exert forces on itself, 
\begin{equation}
\int \frac{d{\bf p}}{(2\pi)^3}
\tilde A_{ij}(\lambda {\bf k}-{\bf p}) = A_{ij}(0) = 0.
\end{equation}
With these relations, one may demonstrate that Eq.~\eqref{eq:phi3} gives no contribution to Eq.~\eqref{eq:stp2}.

The four point correlation function needed for evaluation of Eq.~\eqref{eq:sigsig} is found to be 
\begin{widetext}
\begin{eqnarray}
&&\langle \tilde \rho({\bf k},t) \tilde \rho({\bf k}',t)\tilde \rho({\bf q},0) \tilde \rho({\bf q}',0) \rangle^{\rm id} =  \nonumber \\ 
&& [\overline \rho (2\pi)^3\delta({\bf k} + {\bf k}' +{\bf q}+{\bf q}') {\cal C}({\bf k}+{\bf k}',t)
+(2\pi)^6\overline \rho^2[ \delta({\bf k}+{\bf k}')\delta({\bf q}+{\bf q}')
+ \delta({\bf k}+{\bf q})\delta({\bf k}'+{\bf q}'){\cal C}({\bf k},t){\cal C}({\bf k}',t)+\nonumber \\ &&\delta({\bf k}+{\bf q}')\delta({\bf k}'+{\bf q}){\cal C}({\bf k},t){\cal C}({\bf k}',t) ]\nonumber \\
&+& (2\pi)^6\overline\rho^2[\delta({\bf k}+{\bf k}'+{\bf q})\delta({\bf q}'){\cal C}({\bf k}+{\bf k}',t)+\delta({\bf k}+{\bf k}'+{\bf q}')\delta({\bf q}){\cal C}({\bf k}+{\bf k}',t)+ \delta({\bf k}+{\bf q}+{\bf q}')\delta({\bf k}') {\cal C}({\bf k},t)+ \nonumber \\ &&\delta({\bf k}'+{\bf q}'+{\bf q})\delta({\bf k}){\cal C}({\bf k}',t)]\nonumber\\
&+&(2\pi)^9\overline\rho^3[\delta({\bf k}+{\bf k}')\delta({\bf q})\delta({\bf q}')
+ \delta({\bf k}+{\bf q})\delta({\bf k}')\delta({\bf q}'){\cal C}({\bf k},t)+\delta({\bf k}+{\bf q}')\delta({\bf k}')\delta({\bf q}){\cal C}({\bf k},t)+ \delta({\bf k}'+{\bf q})\delta({\bf k})\delta({\bf q}'){\cal C}({\bf k}',t)+\nonumber \\ && \delta({\bf k}'+{\bf q}')\delta({\bf k})\delta({\bf q}){\cal C}({\bf k}',t)+\delta({\bf q}+{\bf q}')\delta({\bf k}')\delta({\bf k})] + (2\pi)^{12}\overline\rho^4\delta({\bf k})\delta({\bf k}')\delta({\bf q})\delta({\bf q}')]\nonumber
\end{eqnarray}
where we introduced ${\cal C}(\vct{k},t)=e^{-Dk^2t}$.
With this, Eq.~\eqref{eq:phi4} in the main text is found, and Eq.~\eqref{sm}  (for $i\not=j$ and $k\not=l$) via
\begin{equation}
\langle \tilde \sigma_{ij}({\bf k},t)\tilde \sigma_{kl}({\bf q},0)\rangle = \frac{1}{4}\int_0^1 d\lambda d\lambda'\int \frac{d{\bf p}d{\bf  r}}{(2\pi)^6}
\tilde A_{ij}(\lambda {\bf k}-{\bf p})\tilde A_{kl}(\lambda' {\bf q}-{\bf r})\langle \tilde \phi({\bf p},t)\tilde \phi({\bf k}-{\bf p},t)\tilde \phi({\bf r},0)\tilde \phi({\bf q}-{\bf r},0)\rangle.
\end{equation}
\end{widetext}

\section{Effective Hamiltonian of Ref.~\cite{Kruger17}}\label{sec:eH}
The effective Gaussian Hamiltonian for the system of Brownian particles proposed in Ref.~\cite{Kruger17} (see Ref.~\cite{Chandler93} for a partly related approach) is, 
\begin{align}\label{eq:H}
H=\frac{1}{2}\int  d\vct{x} d\vct{y}  \phi(\vct{x}) \Delta(\vct{x},\vct{y})\phi(\vct{y}).
\end{align}
This form contains a quadratic potential which plays the role of an effective interaction potential between densities,
\begin{align}
\Delta(\vct{x},\vct{y})\equiv k_BT\left(\frac{1}{\bar\rho(\vct{x})} \delta(\vct{x}-\vct{y})-c^{(2)}(\vct{x},\vct{y})\label{eq:D}\right).
\end{align}
As discussed in Ref.~\cite{Kruger17}, the Hamiltonian in Eq.~\eqref{eq:H} yields the correct result for the first two moments of the fluctuating field $\phi$ in equilibrium, i.e., $\langle\phi\rangle^{\rm eq}=0$ and 
\begin{align}\label{eq:Ceq}
\langle\phi(\vct{x})\phi(\vct{y})\rangle^{\rm eq}=\left(\frac{1}{\bar\rho(\vct{x})} \delta(\vct{x}-\vct{y})-c^{(2)}(\vct{x},\vct{y})\right)^{-1}. 
\end{align}
The right hand side of Eq.~\eqref{eq:Ceq} is to be understood in the sense of inverse operators. The non-trivial part in the Hamiltonian in Eq.~\eqref{eq:H} is the direct correlation function $c^{(2)}$, which is an important and well-studied object in the theory of liquids \cite{HansenMcDonald}. The corresponding Model B equation of motion reads \cite{Kruger17}
\begin{align}
\notag\frac{\partial \phi}{\partial t}&=\frac{D}{k_BT}\nabla \cdot\left( \langle\rho\rangle\nabla \frac{\delta H}{\delta \phi} \right)+ \nabla \cdot\left(\sqrt{2D\langle\rho\rangle}\boldsymbol\eta\right),\\
&= R\Delta\phi+\nabla \cdot\left(\sqrt{2D\langle\rho\rangle}\boldsymbol\eta\right),\label{eq:eom3}
\end{align}
with noise correlations given in Eq.~\eqref{eq:FDT}. 
With this, Eq.~\eqref{eq:Cm} of the main text is found for small wave-vectors.

\end{appendix}


\begin{thebibliography}{64}%
\makeatletter
\providecommand \@ifxundefined [1]{%
 \@ifx{#1\undefined}
}%
\providecommand \@ifnum [1]{%
 \ifnum #1\expandafter \@firstoftwo
 \else \expandafter \@secondoftwo
 \fi
}%
\providecommand \@ifx [1]{%
 \ifx #1\expandafter \@firstoftwo
 \else \expandafter \@secondoftwo
 \fi
}%
\providecommand \natexlab [1]{#1}%
\providecommand \enquote  [1]{``#1''}%
\providecommand \bibnamefont  [1]{#1}%
\providecommand \bibfnamefont [1]{#1}%
\providecommand \citenamefont [1]{#1}%
\providecommand \href@noop [0]{\@secondoftwo}%
\providecommand \href [0]{\begingroup \@sanitize@url \@href}%
\providecommand \@href[1]{\@@startlink{#1}\@@href}%
\providecommand \@@href[1]{\endgroup#1\@@endlink}%
\providecommand \@sanitize@url [0]{\catcode `\\12\catcode `\$12\catcode
  `\&12\catcode `\#12\catcode `\^12\catcode `\_12\catcode `\%12\relax}%
\providecommand \@@startlink[1]{}%
\providecommand \@@endlink[0]{}%
\providecommand \url  [0]{\begingroup\@sanitize@url \@url }%
\providecommand \@url [1]{\endgroup\@href {#1}{\urlprefix }}%
\providecommand \urlprefix  [0]{URL }%
\providecommand \Eprint [0]{\href }%
\providecommand \doibase [0]{http://dx.doi.org/}%
\providecommand \selectlanguage [0]{\@gobble}%
\providecommand \bibinfo  [0]{\@secondoftwo}%
\providecommand \bibfield  [0]{\@secondoftwo}%
\providecommand \translation [1]{[#1]}%
\providecommand \BibitemOpen [0]{}%
\providecommand \bibitemStop [0]{}%
\providecommand \bibitemNoStop [0]{.\EOS\space}%
\providecommand \EOS [0]{\spacefactor3000\relax}%
\providecommand \BibitemShut  [1]{\csname bibitem#1\endcsname}%
\let\auto@bib@innerbib\@empty
\bibitem [{\citenamefont {Dhont}(1996)}]{dhont}%
  \BibitemOpen
  \bibfield  {author} {\bibinfo {author} {\bibfnamefont {J.~K.~G.}\
  \bibnamefont {Dhont}},\ }\href@noop {} {\emph {\bibinfo {title} {An
  Introduction to Dynamics of Colloids}}}\ (\bibinfo  {publisher} {Elsevier
  science},\ \bibinfo {address} {Amsterdam},\ \bibinfo {year}
  {1996})\BibitemShut {NoStop}%
\bibitem [{\citenamefont {Larson}(1999)}]{larson}%
  \BibitemOpen
  \bibfield  {author} {\bibinfo {author} {\bibfnamefont {R.~G.}\ \bibnamefont
  {Larson}},\ }\href@noop {} {\emph {\bibinfo {title} {The Structure and
  Rheology of Complex Fluids}}}\ (\bibinfo  {publisher} {Oxford University
  Press},\ \bibinfo {address} {New York},\ \bibinfo {year} {1999})\BibitemShut
  {NoStop}%
\bibitem [{\citenamefont {Ortiz~de Z\'arate}\ and\ \citenamefont
  {Sengers}(2006)}]{Ortiz}%
  \BibitemOpen
  \bibfield  {author} {\bibinfo {author} {\bibfnamefont {J.~M.}\ \bibnamefont
  {Ortiz~de Z\'arate}}\ and\ \bibinfo {author} {\bibfnamefont {J.~V.}\
  \bibnamefont {Sengers}},\ }\href@noop {} {\emph {\bibinfo {title}
  {Hydrodynamic Fluctuations in Fluids and Fluid Mixtures}}}\ (\bibinfo
  {publisher} {Elsevier, Amsterdam},\ \bibinfo {year} {2006})\BibitemShut
  {NoStop}%
\bibitem [{\citenamefont {Debenedetti}\ and\ \citenamefont
  {Stillinger}(2001)}]{Debenedetti01}%
  \BibitemOpen
  \bibfield  {author} {\bibinfo {author} {\bibfnamefont {P.~G.}\ \bibnamefont
  {Debenedetti}}\ and\ \bibinfo {author} {\bibfnamefont {F.~H.}\ \bibnamefont
  {Stillinger}},\ }\href@noop {} {\bibfield  {journal} {\bibinfo  {journal}
  {Nature}\ }\textbf {\bibinfo {volume} {410}},\ \bibinfo {pages} {259}
  (\bibinfo {year} {2001})}\BibitemShut {NoStop}%
\bibitem [{\citenamefont {Hansen}\ and\ \citenamefont
  {McDonald}(2009)}]{HansenMcDonald}%
  \BibitemOpen
  \bibfield  {author} {\bibinfo {author} {\bibfnamefont {J.-P.}\ \bibnamefont
  {Hansen}}\ and\ \bibinfo {author} {\bibfnamefont {I.~R.}\ \bibnamefont
  {McDonald}},\ }\href@noop {} {\emph {\bibinfo {title} {Theory of simple
  liquids}}}\ (\bibinfo  {publisher} {Academic Press},\ \bibinfo {year}
  {2009})\BibitemShut {NoStop}%
\bibitem [{\citenamefont {Evans}(1979)}]{bob_advances}%
  \BibitemOpen
  \bibfield  {author} {\bibinfo {author} {\bibfnamefont {R.}~\bibnamefont
  {Evans}},\ }\href@noop {} {\bibfield  {journal} {\bibinfo  {journal}
  {Adv.Phys.}\ }\textbf {\bibinfo {volume} {28}},\ \bibinfo {pages} {143}
  (\bibinfo {year} {1979})}\BibitemShut {NoStop}%
\bibitem [{\citenamefont {Roth}(2010)}]{roth_review}%
  \BibitemOpen
  \bibfield  {author} {\bibinfo {author} {\bibfnamefont {R.}~\bibnamefont
  {Roth}},\ }\href@noop {} {\bibfield  {journal} {\bibinfo  {journal}
  {J.Phys.:Condens.Matter}\ }\textbf {\bibinfo {volume} {22}},\ \bibinfo
  {pages} {063102} (\bibinfo {year} {2010})}\BibitemShut {NoStop}%
\bibitem [{\citenamefont {da~C.~Andrade}(1934)}]{Andrade}%
  \BibitemOpen
  \bibfield  {author} {\bibinfo {author} {\bibfnamefont {E.}~\bibnamefont
  {da~C.~Andrade}},\ }\href@noop {} {\bibfield  {journal} {\bibinfo  {journal}
  {London Edinb. Dub. Philos. Mag. J. Sci.}\ }\textbf {\bibinfo {volume}
  {17(112)}},\ \bibinfo {pages} {497} (\bibinfo {year} {1934})}\BibitemShut
  {NoStop}%
\bibitem [{\citenamefont {Brady}\ and\ \citenamefont
  {Morris}(1997)}]{brady_morris}%
  \BibitemOpen
  \bibfield  {author} {\bibinfo {author} {\bibfnamefont {J.}~\bibnamefont
  {Brady}}\ and\ \bibinfo {author} {\bibfnamefont {J.}~\bibnamefont {Morris}},\
  }\href@noop {} {\bibfield  {journal} {\bibinfo  {journal} {J. Fluid. Mech.}\
  }\textbf {\bibinfo {volume} {348}},\ \bibinfo {pages} {103} (\bibinfo {year}
  {1997})}\BibitemShut {NoStop}%
\bibitem [{\citenamefont {Harris}(2204)}]{Harris04}%
  \BibitemOpen
  \bibfield  {author} {\bibinfo {author} {\bibfnamefont {S.}~\bibnamefont
  {Harris}},\ }\href@noop {} {\emph {\bibinfo {title} {An Introduction to the
  Theory of the Boltzmann Equation}}}\ (\bibinfo  {publisher} {Dover
  Publications, Mineola, New York},\ \bibinfo {year} {2204})\BibitemShut
  {NoStop}%
\bibitem [{\citenamefont {Fuchs}\ and\ \citenamefont {Cates}(2002)}]{Fuchs02}%
  \BibitemOpen
  \bibfield  {author} {\bibinfo {author} {\bibfnamefont {M.}~\bibnamefont
  {Fuchs}}\ and\ \bibinfo {author} {\bibfnamefont {M.~E.}\ \bibnamefont
  {Cates}},\ }\href@noop {} {\bibfield  {journal} {\bibinfo  {journal} {Phys.
  Rev. Lett.}\ }\textbf {\bibinfo {volume} {89}} (\bibinfo {year}
  {2002})}\BibitemShut {NoStop}%
\bibitem [{\citenamefont {Miyazaki}\ and\ \citenamefont
  {Reichman}(2002)}]{Miyazaki02}%
  \BibitemOpen
  \bibfield  {author} {\bibinfo {author} {\bibfnamefont {K.}~\bibnamefont
  {Miyazaki}}\ and\ \bibinfo {author} {\bibfnamefont {D.~R.}\ \bibnamefont
  {Reichman}},\ }\href@noop {} {\bibfield  {journal} {\bibinfo  {journal}
  {Phys. Rev. E}\ }\textbf {\bibinfo {volume} {66}},\ \bibinfo {pages} {050501}
  (\bibinfo {year} {2002})}\BibitemShut {NoStop}%
\bibitem [{\citenamefont {Archer}\ and\ \citenamefont {Evans}(2004)}]{archer}%
  \BibitemOpen
  \bibfield  {author} {\bibinfo {author} {\bibfnamefont {A.~J.}\ \bibnamefont
  {Archer}}\ and\ \bibinfo {author} {\bibfnamefont {R.}~\bibnamefont {Evans}},\
  }\href@noop {} {\bibfield  {journal} {\bibinfo  {journal} {J. Chem. Phys.}\
  }\textbf {\bibinfo {volume} {121}},\ \bibinfo {pages} {4246} (\bibinfo {year}
  {2004})}\BibitemShut {NoStop}%
\bibitem [{\citenamefont {Marconi}\ and\ \citenamefont
  {Tarazona}(1999)}]{Marconi99}%
  \BibitemOpen
  \bibfield  {author} {\bibinfo {author} {\bibfnamefont {U.~M.~B.}\
  \bibnamefont {Marconi}}\ and\ \bibinfo {author} {\bibfnamefont
  {P.}~\bibnamefont {Tarazona}},\ }\href@noop {} {\bibfield  {journal}
  {\bibinfo  {journal} {J. Chem. Phys.}\ }\textbf {\bibinfo {volume} {110}},\
  \bibinfo {pages} {8032} (\bibinfo {year} {1999})}\BibitemShut {NoStop}%
\bibitem [{\citenamefont {Schmidt}\ and\ \citenamefont
  {Brader}(2013)}]{SchmidtBraderJCP_2013_power_func}%
  \BibitemOpen
  \bibfield  {author} {\bibinfo {author} {\bibfnamefont {M.}~\bibnamefont
  {Schmidt}}\ and\ \bibinfo {author} {\bibfnamefont {J.~M.}\ \bibnamefont
  {Brader}},\ }\href@noop {} {\bibfield  {journal} {\bibinfo  {journal} {J.
  Chem. Phys.}\ }\textbf {\bibinfo {volume} {138}},\ \bibinfo {pages} {214101}
  (\bibinfo {year} {2013})}\BibitemShut {NoStop}%
\bibitem [{\citenamefont {Zhang}\ \emph {et~al.}(2004)\citenamefont {Zhang},
  \citenamefont {Todd},\ and\ \citenamefont {Travis}}]{Zhang2004SimPoisFl}%
  \BibitemOpen
  \bibfield  {author} {\bibinfo {author} {\bibfnamefont {J.}~\bibnamefont
  {Zhang}}, \bibinfo {author} {\bibfnamefont {B.~D.}\ \bibnamefont {Todd}}, \
  and\ \bibinfo {author} {\bibfnamefont {K.~P.}\ \bibnamefont {Travis}},\
  }\href@noop {} {\bibfield  {journal} {\bibinfo  {journal} {J. Chem. Phys.}\
  }\textbf {\bibinfo {volume} {121}},\ \bibinfo {pages} {10778} (\bibinfo
  {year} {2004})}\BibitemShut {NoStop}%
\bibitem [{\citenamefont {Glavatskiy}\ \emph {et~al.}(2015)\citenamefont
  {Glavatskiy}, \citenamefont {Dalton}, \citenamefont {Daivis},\ and\
  \citenamefont {Todd}}]{Glavatskiy15}%
  \BibitemOpen
  \bibfield  {author} {\bibinfo {author} {\bibfnamefont {K.}~\bibnamefont
  {Glavatskiy}}, \bibinfo {author} {\bibfnamefont {B.}~\bibnamefont {Dalton}},
  \bibinfo {author} {\bibfnamefont {P.}~\bibnamefont {Daivis}}, \ and\ \bibinfo
  {author} {\bibfnamefont {B.}~\bibnamefont {Todd}},\ }\href@noop {} {\bibfield
   {journal} {\bibinfo  {journal} {Phys. Rev. E}\ }\textbf {\bibinfo {volume}
  {91}},\ \bibinfo {pages} {062132} (\bibinfo {year} {2015})}\BibitemShut
  {NoStop}%
\bibitem [{\citenamefont {Dalton}\ \emph {et~al.}(2015)\citenamefont {Dalton},
  \citenamefont {Glavatskiy}, \citenamefont {Daivis},\ and\ \citenamefont
  {Todd}}]{Glavatskiy15b}%
  \BibitemOpen
  \bibfield  {author} {\bibinfo {author} {\bibfnamefont {B.}~\bibnamefont
  {Dalton}}, \bibinfo {author} {\bibfnamefont {K.}~\bibnamefont {Glavatskiy}},
  \bibinfo {author} {\bibfnamefont {P.}~\bibnamefont {Daivis}}, \ and\ \bibinfo
  {author} {\bibfnamefont {B.}~\bibnamefont {Todd}},\ }\href@noop {} {\bibfield
   {journal} {\bibinfo  {journal} {Ph1ys. Rev. E}\ }\textbf {\bibinfo {volume}
  {92}},\ \bibinfo {pages} {012108} (\bibinfo {year} {2015})}\BibitemShut
  {NoStop}%
\bibitem [{\citenamefont {Roy}\ \emph {et~al.}(2016)\citenamefont {Roy},
  \citenamefont {Dietrich},\ and\ \citenamefont {H{\"o}fling}}]{roy2016}%
  \BibitemOpen
  \bibfield  {author} {\bibinfo {author} {\bibfnamefont {S.}~\bibnamefont
  {Roy}}, \bibinfo {author} {\bibfnamefont {S.}~\bibnamefont {Dietrich}}, \
  and\ \bibinfo {author} {\bibfnamefont {F.}~\bibnamefont {H{\"o}fling}},\
  }\href@noop {} {\bibfield  {journal} {\bibinfo  {journal} {The Journal of
  chemical physics}\ }\textbf {\bibinfo {volume} {145}},\ \bibinfo {pages}
  {134505} (\bibinfo {year} {2016})}\BibitemShut {NoStop}%
\bibitem [{\citenamefont {Hohenberg}\ and\ \citenamefont
  {Halperin}(1977)}]{hohenberg}%
  \BibitemOpen
  \bibfield  {author} {\bibinfo {author} {\bibfnamefont {P.}~\bibnamefont
  {Hohenberg}}\ and\ \bibinfo {author} {\bibfnamefont {B.}~\bibnamefont
  {Halperin}},\ }\href@noop {} {\bibfield  {journal} {\bibinfo  {journal} {Rev.
  Mod. Phys.}\ }\textbf {\bibinfo {volume} {49}} (\bibinfo {year}
  {1977})}\BibitemShut {NoStop}%
\bibitem [{\citenamefont {Onuki}(2002)}]{onukibook}%
  \BibitemOpen
  \bibfield  {author} {\bibinfo {author} {\bibfnamefont {A.}~\bibnamefont
  {Onuki}},\ }\href@noop {} {\emph {\bibinfo {title} {Phase transition
  dynamics}}}\ (\bibinfo  {publisher} {Cambridge University Press},\ \bibinfo
  {year} {2002})\BibitemShut {NoStop}%
\bibitem [{\citenamefont {Kardar}(2007)}]{kardarbook}%
  \BibitemOpen
  \bibfield  {author} {\bibinfo {author} {\bibfnamefont {M.}~\bibnamefont
  {Kardar}},\ }\href@noop {} {\emph {\bibinfo {title} {Statistical physics of
  fields}}}\ (\bibinfo  {publisher} {Cambridge University Press},\ \bibinfo
  {year} {2007})\BibitemShut {NoStop}%
\bibitem [{\citenamefont {Gambassi}\ and\ \citenamefont
  {Dietrich}(2006)}]{gambassi2006}%
  \BibitemOpen
  \bibfield  {author} {\bibinfo {author} {\bibfnamefont {A.}~\bibnamefont
  {Gambassi}}\ and\ \bibinfo {author} {\bibfnamefont {S.}~\bibnamefont
  {Dietrich}},\ }\href@noop {} {\bibfield  {journal} {\bibinfo  {journal} {J.
  Stat. Phys.}\ }\textbf {\bibinfo {volume} {123}} (\bibinfo {year}
  {2006})}\BibitemShut {NoStop}%
\bibitem [{\citenamefont {Gambassi}(2008)}]{gambassi2008EPJB}%
  \BibitemOpen
  \bibfield  {author} {\bibinfo {author} {\bibfnamefont {A.}~\bibnamefont
  {Gambassi}},\ }\href@noop {} {\bibfield  {journal} {\bibinfo  {journal} {Eur.
  Phys. J. B}\ }\textbf {\bibinfo {volume} {64}},\ \bibinfo {pages} {379}
  (\bibinfo {year} {2008})}\BibitemShut {NoStop}%
\bibitem [{\citenamefont {Dean}\ and\ \citenamefont
  {Gopinathan}(2010)}]{deangopinathan2010PRE}%
  \BibitemOpen
  \bibfield  {author} {\bibinfo {author} {\bibfnamefont {D.~S.}\ \bibnamefont
  {Dean}}\ and\ \bibinfo {author} {\bibfnamefont {A.}~\bibnamefont
  {Gopinathan}},\ }\href@noop {} {\bibfield  {journal} {\bibinfo  {journal}
  {Phys. Rev. E}\ }\textbf {\bibinfo {volume} {81}},\ \bibinfo {pages} {041126}
  (\bibinfo {year} {2010})}\BibitemShut {NoStop}%
\bibitem [{\citenamefont {D{\'e}mery}\ and\ \citenamefont
  {Dean}(2010)}]{demery2010}%
  \BibitemOpen
  \bibfield  {author} {\bibinfo {author} {\bibfnamefont {V.}~\bibnamefont
  {D{\'e}mery}}\ and\ \bibinfo {author} {\bibfnamefont {D.~S.}\ \bibnamefont
  {Dean}},\ }\href@noop {} {\bibfield  {journal} {\bibinfo  {journal} {Phys.
  Rev. Lett.}\ }\textbf {\bibinfo {volume} {104}} (\bibinfo {year}
  {2010})}\BibitemShut {NoStop}%
\bibitem [{\citenamefont {Hanke}(2013)}]{hanke}%
  \BibitemOpen
  \bibfield  {author} {\bibinfo {author} {\bibfnamefont {A.}~\bibnamefont
  {Hanke}},\ }\href@noop {} {\bibfield  {journal} {\bibinfo  {journal} {PloS
  {O}ne}\ }\textbf {\bibinfo {volume} {8}} (\bibinfo {year}
  {2013})}\BibitemShut {NoStop}%
\bibitem [{\citenamefont {Furukawa}\ \emph {et~al.}(2013)\citenamefont
  {Furukawa}, \citenamefont {Gambassi}, \citenamefont {Dietrich},\ and\
  \citenamefont {Tanaka}}]{gambassi2013prl}%
  \BibitemOpen
  \bibfield  {author} {\bibinfo {author} {\bibfnamefont {A.}~\bibnamefont
  {Furukawa}}, \bibinfo {author} {\bibfnamefont {A.}~\bibnamefont {Gambassi}},
  \bibinfo {author} {\bibfnamefont {S.}~\bibnamefont {Dietrich}}, \ and\
  \bibinfo {author} {\bibfnamefont {H.}~\bibnamefont {Tanaka}},\ }\href@noop {}
  {\bibfield  {journal} {\bibinfo  {journal} {Phys. Rev. Lett.}\ }\textbf
  {\bibinfo {volume} {111}} (\bibinfo {year} {2013})}\BibitemShut {NoStop}%
\bibitem [{\citenamefont {Rohwer}\ \emph
  {et~al.}(2017{\natexlab{a}})\citenamefont {Rohwer}, \citenamefont
  {Gambassi},\ and\ \citenamefont {Kr\"uger}}]{Rohwer17b}%
  \BibitemOpen
  \bibfield  {author} {\bibinfo {author} {\bibfnamefont {C.~M.}\ \bibnamefont
  {Rohwer}}, \bibinfo {author} {\bibfnamefont {A.}~\bibnamefont {Gambassi}}, \
  and\ \bibinfo {author} {\bibfnamefont {M.}~\bibnamefont {Kr\"uger}},\ }\href
  {http://stacks.iop.org/0953-8984/29/i=33/a=335101} {\bibfield  {journal}
  {\bibinfo  {journal} {Journal of Physics: Condensed Matter}\ }\textbf
  {\bibinfo {volume} {29}},\ \bibinfo {pages} {335101} (\bibinfo {year}
  {2017}{\natexlab{a}})}\BibitemShut {NoStop}%
\bibitem [{\citenamefont {Wada}\ and\ \citenamefont
  {Sasa}(2003)}]{wadasasa2003}%
  \BibitemOpen
  \bibfield  {author} {\bibinfo {author} {\bibfnamefont {H.}~\bibnamefont
  {Wada}}\ and\ \bibinfo {author} {\bibfnamefont {S.-i.}\ \bibnamefont
  {Sasa}},\ }\href {\doibase 10.1103/PhysRevE.67.065302} {\bibfield  {journal}
  {\bibinfo  {journal} {Phys. Rev. E}\ }\textbf {\bibinfo {volume} {67}},\
  \bibinfo {pages} {065302} (\bibinfo {year} {2003})}\BibitemShut {NoStop}%
\bibitem [{\citenamefont {Cattuto}\ \emph {et~al.}(2006)\citenamefont
  {Cattuto}, \citenamefont {Brito}, \citenamefont {Marconi}, \citenamefont
  {Nori},\ and\ \citenamefont {Soto}}]{sotogranular2006}%
  \BibitemOpen
  \bibfield  {author} {\bibinfo {author} {\bibfnamefont {C.}~\bibnamefont
  {Cattuto}}, \bibinfo {author} {\bibfnamefont {R.}~\bibnamefont {Brito}},
  \bibinfo {author} {\bibfnamefont {U.~M.~B.}\ \bibnamefont {Marconi}},
  \bibinfo {author} {\bibfnamefont {F.}~\bibnamefont {Nori}}, \ and\ \bibinfo
  {author} {\bibfnamefont {R.}~\bibnamefont {Soto}},\ }\href {\doibase
  10.1103/PhysRevLett.96.178001} {\bibfield  {journal} {\bibinfo  {journal}
  {Phys. Rev. Lett.}\ }\textbf {\bibinfo {volume} {96}},\ \bibinfo {pages}
  {178001} (\bibinfo {year} {2006})}\BibitemShut {NoStop}%
\bibitem [{\citenamefont {Shaebani}\ \emph {et~al.}(2012)\citenamefont
  {Shaebani}, \citenamefont {Sarabadani},\ and\ \citenamefont
  {Wolf}}]{shaebaniwolf2012}%
  \BibitemOpen
  \bibfield  {author} {\bibinfo {author} {\bibfnamefont {M.~R.}\ \bibnamefont
  {Shaebani}}, \bibinfo {author} {\bibfnamefont {J.}~\bibnamefont
  {Sarabadani}}, \ and\ \bibinfo {author} {\bibfnamefont {D.~E.}\ \bibnamefont
  {Wolf}},\ }\href {\doibase 10.1103/PhysRevLett.108.198001} {\bibfield
  {journal} {\bibinfo  {journal} {Phys. Rev. Lett.}\ }\textbf {\bibinfo
  {volume} {108}},\ \bibinfo {pages} {198001} (\bibinfo {year}
  {2012})}\BibitemShut {NoStop}%
\bibitem [{\citenamefont {Kirkpatrick}\ \emph {et~al.}(2013)\citenamefont
  {Kirkpatrick}, \citenamefont {Ortiz~de Z\'arate},\ and\ \citenamefont
  {Sengers}}]{kirkpatricksengers2013}%
  \BibitemOpen
  \bibfield  {author} {\bibinfo {author} {\bibfnamefont {T.~R.}\ \bibnamefont
  {Kirkpatrick}}, \bibinfo {author} {\bibfnamefont {J.~M.}\ \bibnamefont
  {Ortiz~de Z\'arate}}, \ and\ \bibinfo {author} {\bibfnamefont {J.~V.}\
  \bibnamefont {Sengers}},\ }\href {\doibase 10.1103/PhysRevLett.110.235902}
  {\bibfield  {journal} {\bibinfo  {journal} {Phys. Rev. Lett.}\ }\textbf
  {\bibinfo {volume} {110}},\ \bibinfo {pages} {235902} (\bibinfo {year}
  {2013})}\BibitemShut {NoStop}%
\bibitem [{\citenamefont {Aminov}\ \emph {et~al.}(2015)\citenamefont {Aminov},
  \citenamefont {Kafri},\ and\ \citenamefont {Kardar}}]{aminovkardarkafri2015}%
  \BibitemOpen
  \bibfield  {author} {\bibinfo {author} {\bibfnamefont {A.}~\bibnamefont
  {Aminov}}, \bibinfo {author} {\bibfnamefont {Y.}~\bibnamefont {Kafri}}, \
  and\ \bibinfo {author} {\bibfnamefont {M.}~\bibnamefont {Kardar}},\
  }\href@noop {} {\bibfield  {journal} {\bibinfo  {journal} {Phys. Rev. Lett.}\
  }\textbf {\bibinfo {volume} {114}},\ \bibinfo {pages} {230602} (\bibinfo
  {year} {2015})}\BibitemShut {NoStop}%
\bibitem [{\citenamefont {Kirkpatrick}\ \emph {et~al.}(2015)\citenamefont
  {Kirkpatrick}, \citenamefont {Ortiz~de Z\'arate},\ and\ \citenamefont
  {Sengers}}]{kirkpatrick2015prl}%
  \BibitemOpen
  \bibfield  {author} {\bibinfo {author} {\bibfnamefont {T.~R.}\ \bibnamefont
  {Kirkpatrick}}, \bibinfo {author} {\bibfnamefont {J.~M.}\ \bibnamefont
  {Ortiz~de Z\'arate}}, \ and\ \bibinfo {author} {\bibfnamefont {J.~V.}\
  \bibnamefont {Sengers}},\ }\href {\doibase 10.1103/PhysRevLett.115.035901}
  {\bibfield  {journal} {\bibinfo  {journal} {Phys. Rev. Lett.}\ }\textbf
  {\bibinfo {volume} {115}},\ \bibinfo {pages} {035901} (\bibinfo {year}
  {2015})}\BibitemShut {NoStop}%
\bibitem [{\citenamefont {Kirkpatrick}\ \emph {et~al.}(2016)\citenamefont
  {Kirkpatrick}, \citenamefont {Ortiz~de Z\'arate},\ and\ \citenamefont
  {Sengers}}]{kirkpatrick2016pre}%
  \BibitemOpen
  \bibfield  {author} {\bibinfo {author} {\bibfnamefont {T.~R.}\ \bibnamefont
  {Kirkpatrick}}, \bibinfo {author} {\bibfnamefont {J.~M.}\ \bibnamefont
  {Ortiz~de Z\'arate}}, \ and\ \bibinfo {author} {\bibfnamefont {J.~V.}\
  \bibnamefont {Sengers}},\ }\href {\doibase 10.1103/PhysRevE.93.012148}
  {\bibfield  {journal} {\bibinfo  {journal} {Phys. Rev. E}\ }\textbf {\bibinfo
  {volume} {93}},\ \bibinfo {pages} {012148} (\bibinfo {year}
  {2016})}\BibitemShut {NoStop}%
\bibitem [{\citenamefont {Rohwer}\ \emph
  {et~al.}(2017{\natexlab{b}})\citenamefont {Rohwer}, \citenamefont {Kardar},\
  and\ \citenamefont {Kr\"uger}}]{Rohwer17}%
  \BibitemOpen
  \bibfield  {author} {\bibinfo {author} {\bibfnamefont {C.~M.}\ \bibnamefont
  {Rohwer}}, \bibinfo {author} {\bibfnamefont {M.}~\bibnamefont {Kardar}}, \
  and\ \bibinfo {author} {\bibfnamefont {M.}~\bibnamefont {Kr\"uger}},\
  }\href@noop {} {\bibfield  {journal} {\bibinfo  {journal} {Phys. Rev. Lett.}\
  }\textbf {\bibinfo {volume} {118}},\ \bibinfo {pages} {015702} (\bibinfo
  {year} {2017}{\natexlab{b}})}\BibitemShut {NoStop}%
\bibitem [{\citenamefont {Rohwer}\ \emph {et~al.}()\citenamefont {Rohwer},
  \citenamefont {Solon}, \citenamefont {Kardar},\ and\ \citenamefont
  {Kr\"uger}}]{Rohwer17c}%
  \BibitemOpen
  \bibfield  {author} {\bibinfo {author} {\bibfnamefont {C.~M.}\ \bibnamefont
  {Rohwer}}, \bibinfo {author} {\bibfnamefont {A.}~\bibnamefont {Solon}},
  \bibinfo {author} {\bibfnamefont {M.}~\bibnamefont {Kardar}}, \ and\ \bibinfo
  {author} {\bibfnamefont {M.}~\bibnamefont {Kr\"uger}},\ }\href@noop {}
  {}\bibinfo {note} {ArXiv:1711.11323}\BibitemShut {NoStop}%
\bibitem [{\citenamefont {Solon}\ \emph {et~al.}(2015)\citenamefont {Solon},
  \citenamefont {Fily}, \citenamefont {Baskaran}, \citenamefont {Cates},
  \citenamefont {Kafri}, \citenamefont {Kardar},\ and\ \citenamefont
  {Tailleur}}]{Solon15}%
  \BibitemOpen
  \bibfield  {author} {\bibinfo {author} {\bibfnamefont {A.~P.}\ \bibnamefont
  {Solon}}, \bibinfo {author} {\bibfnamefont {Y.}~\bibnamefont {Fily}},
  \bibinfo {author} {\bibfnamefont {A.}~\bibnamefont {Baskaran}}, \bibinfo
  {author} {\bibfnamefont {M.~E.}\ \bibnamefont {Cates}}, \bibinfo {author}
  {\bibfnamefont {Y.}~\bibnamefont {Kafri}}, \bibinfo {author} {\bibfnamefont
  {M.}~\bibnamefont {Kardar}}, \ and\ \bibinfo {author} {\bibfnamefont
  {J.}~\bibnamefont {Tailleur}},\ }\href@noop {} {\bibfield  {journal}
  {\bibinfo  {journal} {Nature Physics}\ }\textbf {\bibinfo {volume} {11}},\
  \bibinfo {pages} {673} (\bibinfo {year} {2015})}\BibitemShut {NoStop}%
\bibitem [{\citenamefont {Bialk{\'e}}\ \emph {et~al.}(2015)\citenamefont
  {Bialk{\'e}}, \citenamefont {Siebert}, \citenamefont {L{\"o}wen},\ and\
  \citenamefont {Speck}}]{bialke2015negative}%
  \BibitemOpen
  \bibfield  {author} {\bibinfo {author} {\bibfnamefont {J.}~\bibnamefont
  {Bialk{\'e}}}, \bibinfo {author} {\bibfnamefont {J.~T.}\ \bibnamefont
  {Siebert}}, \bibinfo {author} {\bibfnamefont {H.}~\bibnamefont {L{\"o}wen}},
  \ and\ \bibinfo {author} {\bibfnamefont {T.}~\bibnamefont {Speck}},\
  }\href@noop {} {\bibfield  {journal} {\bibinfo  {journal} {Physical review
  letters}\ }\textbf {\bibinfo {volume} {115}},\ \bibinfo {pages} {098301}
  (\bibinfo {year} {2015})}\BibitemShut {NoStop}%
\bibitem [{\citenamefont {Dean}\ and\ \citenamefont
  {Gopinathan}(2009)}]{deangopinathan2009JStatMech}%
  \BibitemOpen
  \bibfield  {author} {\bibinfo {author} {\bibfnamefont {D.~S.}\ \bibnamefont
  {Dean}}\ and\ \bibinfo {author} {\bibfnamefont {A.}~\bibnamefont
  {Gopinathan}},\ }\href@noop {} {\bibfield  {journal} {\bibinfo  {journal} {J.
  Stat. Mech.}\ }\textbf {\bibinfo {volume} {2009}},\ \bibinfo {pages} {L08001}
  (\bibinfo {year} {2009})}\BibitemShut {NoStop}%
\bibitem [{\citenamefont {Bitbol}\ and\ \citenamefont
  {Fournier}(2011)}]{bitbolfournier2011forces}%
  \BibitemOpen
  \bibfield  {author} {\bibinfo {author} {\bibfnamefont {A.-F.}\ \bibnamefont
  {Bitbol}}\ and\ \bibinfo {author} {\bibfnamefont {J.-B.}\ \bibnamefont
  {Fournier}},\ }\href@noop {} {\bibfield  {journal} {\bibinfo  {journal}
  {Physical Review E}\ }\textbf {\bibinfo {volume} {83}},\ \bibinfo {pages}
  {061107} (\bibinfo {year} {2011})}\BibitemShut {NoStop}%
\bibitem [{\citenamefont {Irving}\ and\ \citenamefont
  {Kirkwood}(1950)}]{irving_kirkwood}%
  \BibitemOpen
  \bibfield  {author} {\bibinfo {author} {\bibfnamefont {J.}~\bibnamefont
  {Irving}}\ and\ \bibinfo {author} {\bibfnamefont {J.}~\bibnamefont
  {Kirkwood}},\ }\href@noop {} {\bibfield  {journal} {\bibinfo  {journal} {J.
  Chem. Phys.}\ }\textbf {\bibinfo {volume} {18}},\ \bibinfo {pages} {817}
  (\bibinfo {year} {1950})}\BibitemShut {NoStop}%
\bibitem [{\citenamefont {Kreuzer}(1981)}]{Kreuzer}%
  \BibitemOpen
  \bibfield  {author} {\bibinfo {author} {\bibfnamefont {H.~J.}\ \bibnamefont
  {Kreuzer}},\ }\href@noop {} {\emph {\bibinfo {title} {Nonequilibrium
  thermodynamics and its statistical foundations}}}\ (\bibinfo  {publisher}
  {Clarendon press},\ \bibinfo {address} {Oxford},\ \bibinfo {year}
  {1981})\BibitemShut {NoStop}%
\bibitem [{\citenamefont {Aerov}\ and\ \citenamefont
  {Kr\"uger}(2014)}]{AerovKrugerJCP2014}%
  \BibitemOpen
  \bibfield  {author} {\bibinfo {author} {\bibfnamefont {A.~A.}\ \bibnamefont
  {Aerov}}\ and\ \bibinfo {author} {\bibfnamefont {M.}~\bibnamefont
  {Kr\"uger}},\ }\href@noop {} {\bibfield  {journal} {\bibinfo  {journal} {J.
  Chem. Phys.}\ }\textbf {\bibinfo {volume} {140}},\ \bibinfo {pages} {094701}
  (\bibinfo {year} {2014})}\BibitemShut {NoStop}%
\bibitem [{\citenamefont {Green}(1952)}]{Green}%
  \BibitemOpen
  \bibfield  {author} {\bibinfo {author} {\bibfnamefont {M.}~\bibnamefont
  {Green}},\ }\href@noop {} {\bibfield  {journal} {\bibinfo  {journal} {J.
  Chem. Phys.}\ }\textbf {\bibinfo {volume} {20}},\ \bibinfo {pages} {1281}
  (\bibinfo {year} {1952})}\BibitemShut {NoStop}%
\bibitem [{\citenamefont {Kubo}(1957)}]{Kubo}%
  \BibitemOpen
  \bibfield  {author} {\bibinfo {author} {\bibfnamefont {R.}~\bibnamefont
  {Kubo}},\ }\href@noop {} {\bibfield  {journal} {\bibinfo  {journal} {J. Phys.
  Soc. Jpn.}\ }\textbf {\bibinfo {volume} {12}},\ \bibinfo {pages} {570}
  (\bibinfo {year} {1957})}\BibitemShut {NoStop}%
\bibitem [{\citenamefont {Maier}\ \emph {et~al.}()\citenamefont {Maier},
  \citenamefont {Zippelius},\ and\ \citenamefont {Fuchs}}]{Maier17}%
  \BibitemOpen
  \bibfield  {author} {\bibinfo {author} {\bibfnamefont {M.}~\bibnamefont
  {Maier}}, \bibinfo {author} {\bibfnamefont {A.}~\bibnamefont {Zippelius}}, \
  and\ \bibinfo {author} {\bibfnamefont {M.}~\bibnamefont {Fuchs}},\
  }\href@noop {} {}\bibinfo {note} {ArXiv:1709.09962}\BibitemShut {NoStop}%
\bibitem [{\citenamefont {Evans}(1981)}]{Evans81}%
  \BibitemOpen
  \bibfield  {author} {\bibinfo {author} {\bibfnamefont {D.~J.}\ \bibnamefont
  {Evans}},\ }\href {\doibase 10.1103/PhysRevA.23.2622} {\bibfield  {journal}
  {\bibinfo  {journal} {Phys. Rev. A}\ }\textbf {\bibinfo {volume} {23}},\
  \bibinfo {pages} {2622} (\bibinfo {year} {1981})}\BibitemShut {NoStop}%
\bibitem [{\citenamefont {Aerov}\ and\ \citenamefont
  {Kr{\"u}ger}(2015)}]{aerov2015}%
  \BibitemOpen
  \bibfield  {author} {\bibinfo {author} {\bibfnamefont {A.~A.}\ \bibnamefont
  {Aerov}}\ and\ \bibinfo {author} {\bibfnamefont {M.}~\bibnamefont
  {Kr{\"u}ger}},\ }\href@noop {} {\bibfield  {journal} {\bibinfo  {journal}
  {Physical Review E}\ }\textbf {\bibinfo {volume} {92}},\ \bibinfo {pages}
  {042301} (\bibinfo {year} {2015})}\BibitemShut {NoStop}%
\bibitem [{\citenamefont {Machta}\ \emph {et~al.}(1979)\citenamefont {Machta},
  \citenamefont {Oppenheim},\ and\ \citenamefont {Procaccia}}]{Machta79}%
  \BibitemOpen
  \bibfield  {author} {\bibinfo {author} {\bibfnamefont {J.}~\bibnamefont
  {Machta}}, \bibinfo {author} {\bibfnamefont {I.}~\bibnamefont {Oppenheim}}, \
  and\ \bibinfo {author} {\bibfnamefont {I.}~\bibnamefont {Procaccia}},\
  }\href@noop {} {\bibfield  {journal} {\bibinfo  {journal} {Phys. Rev. Lett.}\
  }\textbf {\bibinfo {volume} {42}},\ \bibinfo {pages} {1368} (\bibinfo {year}
  {1979})}\BibitemShut {NoStop}%
\bibitem [{\citenamefont {Risken}(1984)}]{Risken}%
  \BibitemOpen
  \bibfield  {author} {\bibinfo {author} {\bibfnamefont {H.}~\bibnamefont
  {Risken}},\ }\href@noop {} {\emph {\bibinfo {title} {The Fokker-Planck
  Equation}}}\ (\bibinfo  {publisher} {Springer},\ \bibinfo {address}
  {Berlin},\ \bibinfo {year} {1984})\BibitemShut {NoStop}%
\bibitem [{\citenamefont {Dean}(1996)}]{Dean96}%
  \BibitemOpen
  \bibfield  {author} {\bibinfo {author} {\bibfnamefont {D.~S.}\ \bibnamefont
  {Dean}},\ }\href {http://stacks.iop.org/0305-4470/29/i=24/a=001} {\bibfield
  {journal} {\bibinfo  {journal} {Journal of Physics A: Mathematical and
  General}\ }\textbf {\bibinfo {volume} {29}},\ \bibinfo {pages} {L613}
  (\bibinfo {year} {1996})}\BibitemShut {NoStop}%
\bibitem [{\citenamefont {Wajnryb}\ \emph {et~al.}()\citenamefont {Wajnryb},
  \citenamefont {Altenberger},\ and\ \citenamefont
  {Dahler}}]{wajnryb_uniqueness_1995}%
  \BibitemOpen
  \bibfield  {author} {\bibinfo {author} {\bibfnamefont {E.}~\bibnamefont
  {Wajnryb}}, \bibinfo {author} {\bibfnamefont {A.~R.}\ \bibnamefont
  {Altenberger}}, \ and\ \bibinfo {author} {\bibfnamefont {J.~S.}\ \bibnamefont
  {Dahler}},\ }\href@noop {} {\ \textbf {\bibinfo {volume} {103}},\ \bibinfo
  {pages} {9782}}\BibitemShut {NoStop}%
\bibitem [{\citenamefont {Kr{\"u}ger}\ and\ \citenamefont
  {Dean}(2017)}]{Kruger17}%
  \BibitemOpen
  \bibfield  {author} {\bibinfo {author} {\bibfnamefont {M.}~\bibnamefont
  {Kr{\"u}ger}}\ and\ \bibinfo {author} {\bibfnamefont {D.~S.}\ \bibnamefont
  {Dean}},\ }\href@noop {} {\bibfield  {journal} {\bibinfo  {journal} {The
  Journal of Chemical Physics}\ }\textbf {\bibinfo {volume} {146}},\ \bibinfo
  {pages} {134507} (\bibinfo {year} {2017})}\BibitemShut {NoStop}%
\bibitem [{\citenamefont {Dean}\ and\ \citenamefont
  {Podgornik}(2014)}]{deanpodgornik2014}%
  \BibitemOpen
  \bibfield  {author} {\bibinfo {author} {\bibfnamefont {D.~S.}\ \bibnamefont
  {Dean}}\ and\ \bibinfo {author} {\bibfnamefont {R.}~\bibnamefont
  {Podgornik}},\ }\href@noop {} {\bibfield  {journal} {\bibinfo  {journal}
  {Phys. Rev. E}\ }\textbf {\bibinfo {volume} {89}},\ \bibinfo {pages} {032117}
  (\bibinfo {year} {2014})}\BibitemShut {NoStop}%
\bibitem [{\citenamefont {Lu}\ \emph {et~al.}(2015)\citenamefont {Lu},
  \citenamefont {Dean},\ and\ \citenamefont {Podgornik}}]{lu15}%
  \BibitemOpen
  \bibfield  {author} {\bibinfo {author} {\bibfnamefont {B.-S.}\ \bibnamefont
  {Lu}}, \bibinfo {author} {\bibfnamefont {D.~S.}\ \bibnamefont {Dean}}, \ and\
  \bibinfo {author} {\bibfnamefont {R.}~\bibnamefont {Podgornik}},\ }\href@noop
  {} {\bibfield  {journal} {\bibinfo  {journal} {Europhys. Lett.}\ }\textbf
  {\bibinfo {volume} {112}},\ \bibinfo {pages} {20001} (\bibinfo {year}
  {2015})}\BibitemShut {NoStop}%
\bibitem [{\citenamefont {Dean}\ \emph {et~al.}(2016)\citenamefont {Dean},
  \citenamefont {Lu}, \citenamefont {Maggs},\ and\ \citenamefont
  {Podgornik}}]{dean16}%
  \BibitemOpen
  \bibfield  {author} {\bibinfo {author} {\bibfnamefont {D.~S.}\ \bibnamefont
  {Dean}}, \bibinfo {author} {\bibfnamefont {B.-S.}\ \bibnamefont {Lu}},
  \bibinfo {author} {\bibfnamefont {A.~C.}\ \bibnamefont {Maggs}}, \ and\
  \bibinfo {author} {\bibfnamefont {R.}~\bibnamefont {Podgornik}},\ }\href@noop
  {} {\bibfield  {journal} {\bibinfo  {journal} {Phys. Rev. Lett.}\ }\textbf
  {\bibinfo {volume} {116}},\ \bibinfo {pages} {240602} (\bibinfo {year}
  {2016})}\BibitemShut {NoStop}%
\bibitem [{\citenamefont {D\'emery}\ and\ \citenamefont
  {Dean}(2016)}]{demery16}%
  \BibitemOpen
  \bibfield  {author} {\bibinfo {author} {\bibfnamefont {V.}~\bibnamefont
  {D\'emery}}\ and\ \bibinfo {author} {\bibfnamefont {D.~S.}\ \bibnamefont
  {Dean}},\ }\href@noop {} {\bibfield  {journal} {\bibinfo  {journal} {J. Stat.
  Mech.}\ ,\ \bibinfo {pages} {023106}} (\bibinfo {year} {2016})}\BibitemShut
  {NoStop}%
\bibitem [{\citenamefont {Gibbs}(1902)}]{Gibbs02}%
  \BibitemOpen
  \bibfield  {author} {\bibinfo {author} {\bibfnamefont {J.~W.}\ \bibnamefont
  {Gibbs}},\ }\href@noop {} {\emph {\bibinfo {title} {Elementary principles in
  statistical mechanics}}}\ (\bibinfo  {publisher} {Yale University Press},\
  \bibinfo {year} {1902})\BibitemShut {NoStop}%
\bibitem [{\citenamefont {Velenich}\ \emph {et~al.}(2008)\citenamefont
  {Velenich}, \citenamefont {Chamon}, \citenamefont {Cugliandolo},\ and\
  \citenamefont {Kreimer}}]{velenich08}%
  \BibitemOpen
  \bibfield  {author} {\bibinfo {author} {\bibfnamefont {V.}~\bibnamefont
  {Velenich}}, \bibinfo {author} {\bibfnamefont {C.}~\bibnamefont {Chamon}},
  \bibinfo {author} {\bibfnamefont {L.~F.}\ \bibnamefont {Cugliandolo}}, \ and\
  \bibinfo {author} {\bibfnamefont {D.}~\bibnamefont {Kreimer}},\ }\href@noop
  {} {\bibfield  {journal} {\bibinfo  {journal} {J. Phys. A.}\ }\textbf
  {\bibinfo {volume} {41}},\ \bibinfo {pages} {235002} (\bibinfo {year}
  {2008})}\BibitemShut {NoStop}%
\bibitem [{\citenamefont {Kubo}\ \emph {et~al.}(2012)\citenamefont {Kubo},
  \citenamefont {Toda},\ and\ \citenamefont {Hashitsume}}]{kubobook}%
  \BibitemOpen
  \bibfield  {author} {\bibinfo {author} {\bibfnamefont {R.}~\bibnamefont
  {Kubo}}, \bibinfo {author} {\bibfnamefont {M.}~\bibnamefont {Toda}}, \ and\
  \bibinfo {author} {\bibfnamefont {N.}~\bibnamefont {Hashitsume}},\
  }\href@noop {} {\emph {\bibinfo {title} {{Statistical physics II:
  nonequilibrium statistical mechanics}}}},\ Vol.~\bibinfo {volume} {31}\
  (\bibinfo  {publisher} {Springer Science \& Business Media},\ \bibinfo {year}
  {2012})\BibitemShut {NoStop}%
\bibitem [{\citenamefont {Percus}(1996)}]{percuspaper}%
  \BibitemOpen
  \bibfield  {author} {\bibinfo {author} {\bibfnamefont {J.~K.}\ \bibnamefont
  {Percus}},\ }\href@noop {} {\bibfield  {journal} {\bibinfo  {journal} {J.
  Math. Phys.}\ }\textbf {\bibinfo {volume} {37}},\ \bibinfo {pages} {1259}
  (\bibinfo {year} {1996})}\BibitemShut {NoStop}%
\bibitem [{\citenamefont {Chandler}(1993)}]{Chandler93}%
  \BibitemOpen
  \bibfield  {author} {\bibinfo {author} {\bibfnamefont {D.}~\bibnamefont
  {Chandler}},\ }\href@noop {} {\bibfield  {journal} {\bibinfo  {journal}
  {Phys. Rev. E}\ }\textbf {\bibinfo {volume} {48}},\ \bibinfo {pages} {2898}
  (\bibinfo {year} {1993})}\BibitemShut {NoStop}%
\end{thebibliography}

%

\end{document}